\documentclass[letterpaper,twocolumn,10pt]{article}
\usepackage{usenix-2020-09}

\usepackage[normalem]{ulem}
\usepackage{etex}
\usepackage{tikz}
\usepackage{amsmath}
\usepackage{filecontents}
\usepackage{color}
\usepackage{setspace}
\usepackage[compact]{titlesec}
\usepackage{xspace}
\usepackage{enumitem}
\usepackage{mfirstuc}
\usepackage{caption}
\usepackage[labelformat=simple]{subcaption}
\usepackage{circledsteps}
\usepackage{booktabs}
\usepackage{multirow}
\usepackage{pifont}
\usepackage{tablefootnote}
\usepackage{makecell}
\usepackage{bookmark}
\usepackage[table]{xcolor}
\usepackage{cleveref}
\usepackage{hyperref}
\usepackage[misc]{ifsym}
\usepackage{bbding}
\usepackage{breakurl}
\usepackage{algorithm}
\usepackage{algpseudocode}
\usepackage{listings}
\usepackage[many]{tcolorbox}
\usepackage{listings}
\usepackage[]{mfirstuc}
\usepackage{textcomp}
\hypersetup{
  colorlinks,
  linkcolor={red!50!black},
  citecolor={blue!50!black},
  urlcolor={blue!80!black}
}

\titleformat*{\section}{\large\bfseries}
\titleformat*{\subsection}{\normalsize\bfseries}
\titleformat*{\subsubsection}{\normalsize\bfseries}

\makeatletter
\renewcommand{\sectionautorefname}{\S\@gobble}
\renewcommand{\subsectionautorefname}{\S\@gobble}
\renewcommand{\subsubsectionautorefname}{\S\@gobble}

\setenumerate[1]{itemsep=0pt,partopsep=0pt,parsep=\parskip,topsep=1pt,leftmargin=1em}
\setitemize[1]{itemsep=0pt,partopsep=0pt,parsep=\parskip,topsep=1pt,leftmargin=1em}

\renewcommand{\sectionautorefname}{§\!}
\renewcommand{\subsectionautorefname}{§\!}

\newcommand{\darkred}[1]{{\color{red!70!black}#1}}
\newcommand{\darkblue}[1]{{\color{blue!70!black}#1}}

\newcommand{\emphbf}[1]{{\emph{\textbf{#1}}}}

\newcommand{\myding}[1]{\raisebox{-0.15ex}{\ding{#1}}}

\newcommand{\redding}[1]{\darkred{\myding{#1}}}
\newcommand{\circled}[1]{\raisebox{0.15ex}{\tikz[baseline=(char.base)]{
  \node[shape=circle,draw,inner sep=0.3pt] (char) {#1};}}}
\newcommand{\bluefilledcircled}[1]{\raisebox{0.15ex}{\tikz[baseline=(char.base)]{
  \node[shape=circle,inner sep=0.3pt,fill=blue!70!black,text=white,] (char) {#1};}}}

\tcbset{
    sharp corners,
    colback = white
}
\newcommand{\myboxA}[1]{%
    \vspace{2pt}
    \noindent
    \begin{tcolorbox}[colframe=white, colback=white, arc=1mm, boxsep=0em, left=0.5em,right=0.5em, top=2pt,bottom=2pt, boxrule=0pt,frame hidden,sharp corners,enhanced,borderline west={1.5pt}{0pt}{black}
      ]
        #1
    \end{tcolorbox}
    \vspace{-3px}
  }
\newcommand{\myboxB}[1]{%
    \vspace{3pt}
    \noindent
    \begin{tcolorbox}[colframe=black, colback=white, arc=3mm, boxsep=0em, left=0.5em,right=0.5em, top=3pt,bottom=3pt]
        #1
    \end{tcolorbox}
  }
\newcommand{\guideline}{P$^3$\xspace}
\newcommand{\ruleprefix}{SP\xspace}
\newcommand{\glshort}{SP$^4$\xspace}
\newcommand{\rules}{\ruleprefix guidelines\xspace}
\newcommand{\Rules}{\ruleprefix Guidelines\xspace}
\newcommand{\gl}{{\guideline} guideline\xspace}

\newcommand{\glprefixu}{Out-of-\underline{P}lace update, Re\underline{P}licated shared variable, S\underline{P}eculative Reading\xspace}

\newcommand{\pcc}{partial cache-coherence\xspace}

\newcommand{\ds}{PCCIndex\xspace}

\newcommand{\sysbwtree}{{\guideline}-BwTree\xspace}
\newcommand{\syshash}{{\guideline}-CLevelHash\xspace}
\newcommand{\pccbwtree}{SP-BwTree\xspace}
\newcommand{\pcchash}{SP-CLevelHash\xspace}

\newcommand{\pcas}{\texttt{pCAS}\xspace}
\newcommand{\pload}{\texttt{pLoad}\xspace}
\newcommand{\pstore}{\texttt{pStore}\xspace}

\begin{document}

\title{Guidelines for Building Indexes on Partially Cache-Coherent CXL Shared Memory}

\author{\rm Fangnuo Wu,\; Mingkai Dong,\; Wenjun Cai,\; Jingsheng Yan,\; and Haibo Chen\\
  Institute of Parallel and Distributed Systems (IPADS), Shanghai Jiao Tong University
} %

\maketitle
\pagestyle{plain}

\begin{abstract}

\noindent

The \emph{Partial Cache-Coherence (PCC)} model maintains hardware cache coherence only within subsets of cores, enabling large-scale memory sharing with emerging memory interconnect technologies like Compute Express Link (CXL).
However, PCC's relaxation of global cache coherence compromises the correctness of existing single-machine software.

This paper focuses on building consistent and efficient indexes on PCC platforms.
We present that existing indexes designed for cache-coherent platforms can be made consistent on PCC platforms following {\rules}, i.e., we identify \emph{\underline{S}ync-data} and \emph{\underline{P}rotected-data} according to the index's concurrency control mechanisms, and synchronize them accordingly.
However, conversion with \rules introduces performance overhead.
To mitigate the overhead, we identify several unique performance bottlenecks on PCC platforms, and propose {\gl}s (i.e., using \glprefixu) to improve the efficiency of converted indexes on PCC platforms.

With SP and {\guideline} guidelines, we convert and optimize two indexes (CLevelHash and BwTree) for PCC platforms.
    Evaluation shows that converted indexes' throughput improves up to 16$\times$ following {\gl}s, and the optimized indexes outperform their message-passing-based and disaggregated-memory-based counterparts by up to 16$\times$ and 19$\times$.

\end{abstract}

\section{Introduction}%
\label{sec:intro}

Recent advances in highly efficient and byte-addressable memory interconnect technologies, such as Compute Express Link (CXL)~\cite{web:cxl-spec} and Unified Bus~\cite{web:huawei-ub-apnet21}, enable memory sharing across multiple hosts. 
Cross-host memory sharing enables zero-copy data transfer and shared-everything data access, which benefits various application scenarios including key-value stores\cite{redis-home-page,ramcloud}, database systems\cite{polardb} and distributed computing frameworks\cite{Zaharia2010SparkCC,ray2018}.

However, it's impractical to maintain global hardware cache-coherence for memory shared across large-scale hosts.
While CXL 3.0~\cite{cxl-3.1-spec} has announced hardware cache-coherence support, researchers and CXL vendors show that large-scale hardware cache-coherence introduces non-negligible overhead and fails to scale to a larger number of hosts~\cite{wang2025enabling,jain2024AMD-cxl}.

To enable large-scale memory sharing, one practical approach is to relax global cache-coherence with the \emph{\pcc (PCC) models}~\cite{fu2015coherence,lotfi-kamran2012scale-out}. 
A typical implementation of the PCC model maintains cache coherence among cores within a single host, but not across hosts.
To enable synchronization on PCC platforms, global cache-bypass atomic instructions are essential, and atomics are feasible to be implemented without the significant cost and complexity associated with hardware cache coherence~\cite{zhang2023partial,web:huawei-ub-apnet21}.

This paper focuses on building consistent and efficient indexes on PCC platforms, as indexes are fundamental components of modern software systems~\cite{redis-home-page,memcached-home-page,polardb,ray2018}.
We begin by converting indexes designed for cache-coherent platforms (CCIndex) to PCC platforms (\ds) with \rules.
CCIndexes are selected as they are time-tested and have a variety of implementations.
Then, we analyze the performance bottlenecks of \ds and mitigate them accordingly.

To convert CCIndexes for correct execution on PCC platforms that lack cache coherence, it is essential to correctly control the update visibility of shared data (with their state cached in CPU cache). 
To achieve this, we distinguish between two types of shared data that have different visibility requirements. 
First, \emph{sync-data} used for concurrency control (e.g., lock flag) requires immediate visibility.
Second, \emph{protected-data} protected by synchronization mechanisms (e.g., data protected by locks) can delay visibility, since the concurrency control mechanisms protect it from being accessed by other cores.
We thus propose the \rules to ensure correctness of \ds, which synchronize \emph{sync-data} using cache-bypass atomics for immediate visibility, and use cacheline flush to make \emph{protected-data} visible according to index's concurrency control mechanism.

However, conversion with \rules introduces overhead due to excessive software management of cachelines.
By breaking down the overhead of the converted indexes, we reveal several common bottlenecks for PCC platforms, and propose {\gl}s to mitigate the overhead accordingly.

First, we observe that \emph{in-place updates} of converted \ds introduce more overhead compared with indexes using \emph{out-of-place updates}.
This is because out-of-place updates eliminate the need for invalidating \emph{protected-data}'s cacheline before reading, since \emph{protected-data}'s cacheline is always fresh as out-of-place updates do not modify them in-place.
Besides, out-of-place updated indexes require less cache-bypass operations on \emph{sync-data} compared to in-place updated ones. 
With this observation, we start building \ds based on indexes with \emph{\underline{O}ut-of-place updates}.

Second, we find that current loads, which are commonly considered as scalable on cache-coherent platforms, are actually not scalable on PCC platforms.
The underlying reason is that concurrent cache-bypass loads of the same physical address are strictly ordered by the hardware memory controller to preserve memory semantics~\cite{intel-manual,midhul2024understanding}.
Thus, \ds's scalability is bottlenecked by loads of frequently-accessed shared variables (e.g., root node pointer for tree indexes).
To mitigate this, we use \emph{\underline{R}eplicated shared variables}, which replace cache-bypass loads to a single physical address with loads to multiple distinct addresses.

Third, we find that converted \ds fail to fully exploit the high-speed CPU cache and cache coherence within the host.
This limitation becomes particularly significant under read-heavy and skewed workloads, which are common in real-world production systems~\cite{berk2012workload-analysis,nishtala2013scaling,Yang2020TwitterTrace}.
The primary reason is that converted indexes perform cache-bypass loads even when the data is already cached and unchanged---a common scenario in read-heavy and skewed workloads.
To better leverage the CPU cache, we employ \emph{sp\underline{E}culative reading} instead of always performing cache-bypass loads.
Data can first be speculatively read from the cache; its freshness is then verified, and the operation is retried if necessary to ensure correctness.

Following the \glshort guidelines, we convert and optimize a hash index (CLevel-Hash~\cite{chen2020CLevel}), a B+Tree index (BwTree~\cite{Levandoski2013BwTree}), and a general decentralized garbage collection mechanisms~\cite{tu2013silo,wang2018openbwtree} for PCC platforms.
In the evaluation using YCSB~\cite{YCSB2010} and real-world traces (i.e., Twitter~\cite{Yang2020TwitterTrace}),
applying {\gl}s can improve the throughput of a converted but not optimized index by up to 2.4$\times$ (for CLevelHash) and 16$\times$ (for BwTree).
Besides, \syshash and \sysbwtree achieve up to 16$\times$ and 13$\times$ throughput improvement over partitioned and message-passing-based indexes, and \sysbwtree achieves up to 19$\times$ improvement over Sherman, an RDMA-based index ported to the CXL platform.
Besides, integrating \sysbwtree with Ray results in a 49\% throughput improvement for an RL algorithm~\cite{impala}.

In summary, the paper makes the following contributions.
\begin{itemize}[leftmargin=1em]
\setlength\itemsep{0em}
\item We introduced \rules for converting CCIndex with different synchronization mechanisms to PCCIndex.
\item We revealed the unique inefficiencies of PCC platforms and proposed the {\gl} for unleashing the performance of PCCIndex.
\item We implemented two indexes and a general GC method using \rules and {\gl}s and evaluated their performance using YCSB and real-world traces.
\end{itemize}

\section{Partial Cache-Coherence Shared Memory}%
\label{sec:back}

\subsection{Memory interconnect technologies}
Recent memory interconnect technologies, such as Compute Express Link (CXL)\cite{web:cxl-spec} and Unified Bus\cite{web:huawei-ub-apnet21}, enable multiple physically connected hosts to share memory.
Compared to traditional RDMA-based memory access methods, these technologies offer two key advantages. 
First, they support CPU-cacheable memory semantics (i.e., load and store operations), whereas RDMA-based accesses are not CPU-cacheable and rely on send/receive verbs. 
Second, they provide substantially lower access latency: a cache-hit load/store can complete in only 10--20\,ns, while a cache miss to ASIC/FPGA-based CXL-attached memory via a single CXL switch incurs approximately 300--500\,ns~\cite{wang2025enabling,li2023pond}. In contrast, RDMA-based memory accesses typically exhibit microsecond-level latency.

\subsection{Large-scale cache-coherence is not practical}

\begin{figure}[t]
    \centering
    \includegraphics[width=0.85\linewidth]{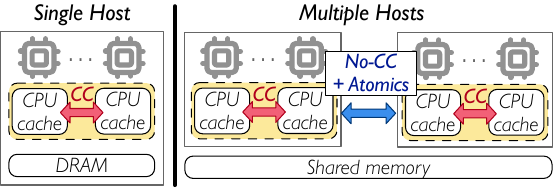}
    \vspace{-8px}
    \caption{\textbf{Partial cache-coherence model.} 
    }
    \label{fig:hardware-model}
\vspace{-12px}
\end{figure}

Achieving cache coherence across large-scale hosts is considered impractical due to the fundamental limitations of directory-based and snooping-based protocols~\cite{Gupta1990ReducingMA,agrawal1988evaluation,xu2011composite}. 
Directory-based approaches incur substantial memory overhead to maintain coherence states, while snooping-based methods require broadcasting messages to all hosts, resulting in excessive communication costs as the system scales~\cite{agrawal1988evaluation}.

Although memory interconnect technologies such as CXL 3.0~\cite{cxl-3.1-spec} claim to support global hardware cache coherence, both academic studies and industry reports show that such support is inefficient and only practical for small-scale systems (e.g., up to 8 hosts)\cite{wang2025enabling,jain2024AMD-cxl}. 
In practice, maintaining hardware cache coherence in CXL 3.0 requires endpoints to track cacheline ownership using snoop filters (SF) and to issue back-invalidate snoop (BISnp) messages to all hosts for cache state updates\cite{cxl-3.1-spec}. 
Even in a simple two-host configuration, this process introduces 44--85\% higher memory access latency~\cite{wang2025enabling}, and the limited capacity of SF leads to frequent conflicts and cacheline evictions, exacerbating performance degradation.
Moreover, as the number of hosts increases, the system must broadcast BISnp messages to all participants and wait for their acknowledgments.
This broadcasting and waiting cause significant delays that are unacceptable for hardware, making hardware cache coherence infeasible for large-scale CXL deployments.

\begin{figure}[t]
    \centering
    \subfloat[Inconsistent \texttt{Load/Store}]{
        \includegraphics[width=0.73\linewidth]{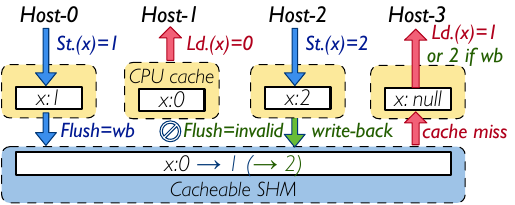}
        \vspace{-4pt}
        \label{fig:pcc-model-rw}
    }
    \subfloat[Atomics]{
        \hspace{-12pt} \includegraphics[width=0.225\linewidth]{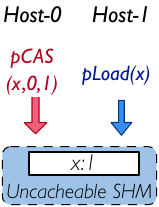}
        \vspace{-4pt}
        \label{fig:pcc-model-cas}
    }
    \vspace{-8px}
    \caption{\textbf{Operations' behavior on PCC platforms.}
   }
    \label{fig:pcc-model-behavior}
    \vspace{-18px}
\end{figure}

\begin{figure*}[t]
    \centering
    \subfloat[CCIndex]{
        \includegraphics[width=0.175\linewidth]{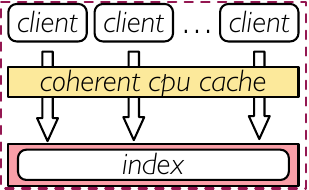}
        \vspace{-5pt}
        \label{fig:ccindex}
    }
    \subfloat[Partitioned Index]{
        \includegraphics[width=0.17\linewidth]{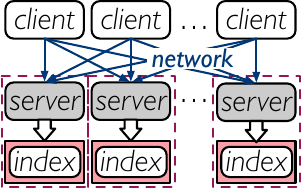}
        \vspace{-5pt}
        \label{fig:index-distributed}
    }
    \subfloat[DM Index]{
        \includegraphics[width=0.17\linewidth]{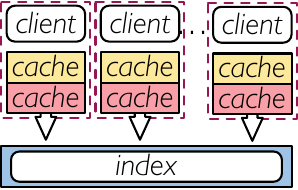}
        \vspace{-5pt}
        \label{fig:index-dm}
    }
    \subfloat[Partitioned (CXL)]{
        \includegraphics[width=0.17\linewidth]{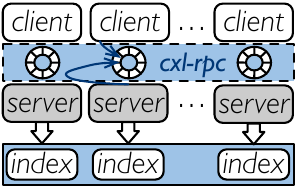}
        \vspace{-5pt}
        \label{fig:index-distributed-cxl}
    }
    \subfloat[PCCIndex (Our)]{
        \includegraphics[width=0.17\linewidth]{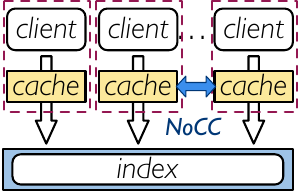}
        \vspace{-5pt}
        \label{fig:pccindex}
    }
    \vspace{-10pt}
    \caption{\textbf{Existing indexes.}
    (a), (c) and (e) employs a share-everything architecture, while (b) and (d) are share-nothing.
    }
    \label{fig:moti-share-everything-nothing}
\vspace{-10px}
\end{figure*}

\subsection{Partial cache-coherence is practical}
\label{sec:moti-pcc-model}
Although maintaining large-scale cache-coherence faces challenges, it is practical to support only partial cache-coherence among large-scale hosts.
We refer to the setup that relaxes cache-coherence across multiple hosts as the \emph{Partial Cache-Coherence (PCC)} model.
One simple setup is for each host to maintain cache coherence among its own CPU cores, but not with other hosts (see \autoref{fig:hardware-model}).
Without cross-host cache coherence, the system can scale independently with the number of hosts, making large-scale setups possible.

To facilitate synchronization across hosts, the PCC platform should support global hardware atomic instructions.
\begin{itemize}
    \item \emphbf{essential}: atomic instructions are essential for implementing synchronization mechanisms such as locks and lock-free schemes. Without atomic instructions, synchronization would have to rely on highly inefficient software-based approaches like software locks~\cite{lamport1974bakery}.
    \item \emphbf{feasible}: atomic instructions are feasible to implement without relying on global cache coherence.
    Existing works and our discussions with CXL vendors~\cite{zhang2023partial} confirm their support of global atomic instructions.
    As shown in \autoref{fig:pcc-model-cas}, \texttt{CAS} and \texttt{Load} can bypass the cache and directly operate on the shared memory, so that the cache coherence protocol is not involved.
    We use \texttt{pCAS} and \texttt{pLoad} to refer to cache-bypass \texttt{CAS} and \texttt{Load}, respectively.
    \texttt{pLoad} is already supported on existing platforms by using non-temporal load instructions or calling \texttt{clflush/mfence} + \texttt{Load}.
    In contrast, \texttt{pCAS} is not supported on existing platforms, and can be implemented by sending MMIO-based requests and memory controller can execute \texttt{pCAS} requests atomically. 
    
\end{itemize}

\subsection{Inconsistency caused by PCC model} 
The PCC model introduces inconsistent CPU caches across multiple hosts, which causes issues for applications that rely on cache-coherence.
First, each host might \texttt{Load} stale data.
As shown in \autoref{fig:pcc-model-rw}, Host-1 \texttt{Load}s a cached value of $x=0$, while $x$ on shared memory is already updated to 1 by Host-0.
Second, \texttt{Store} is not immediately visible to other hosts.
For example, Host-0 and Host-2's \texttt{Store}s are cached in their CPU cache, and are visible to other hosts only after cacheline flush (Host-0's store is flushed thus visible, but Host-2's stores are still hidden).
Third, cache agents might trigger a cacheline \emph{write-back} at any time.
For example, Host-2's cached data ($x=2$) might be written back to shared memory before Host-2 explicitly calls cacheline flush instructions.

\section{PCC Indexing: Motivation and Challenges}

\subsection{Motivation: supporting indexes on PCC platforms}

Indexes are pivotal components of a wide range of applications (e.g., key-value stores~\cite{redis-home-page,memcached-home-page}, databases~\cite{polardb,ramcloud}, distributed processing frameworks~\cite{ray2018,Murray2011ciel,dis-pytorch}).
Memory interconnect technologies enable efficient cross-host in-memory indexes, which open up opportunities for improving the performance of these applications.
For example, Ray~\cite{ray2018}, a distributed AI/ML framework, utilizes indexes as object stores to communicate among multiple workers on different hosts. 
With an efficient cross-host index, Ray's workers can directly access objects on shared memory, which is much more efficient than network-based communication since it avoids data copy, serialization and data transfer over the network.

\subsection{Existing indexes and limitations on PCC platforms}

By reviewing existing indexes, we find that they are either incorrect or inefficient on PCC platforms.

\emphbf{CCIndexes (\autoref{fig:ccindex}) are incorrect.}
CCIndexes are designed for cache-coherent shared memory.
They often use locks/lock-free techniques for concurrency control.
These techniques heavily rely on global cache coherence and cannot function correctly on PCC platforms.

\emphbf{Distributed partitioned indexes (\autoref{fig:index-distributed}, \autoref{fig:index-distributed-cxl}) are inefficient.}
Distributed indexes consist of partitioned data structures on each host.
Readers/writers use message passing to send requests to the host owning the corresponding partition to access the data.
Distributed indexes can run correctly on PCC platforms, but their shared-nothing architecture leads to inefficiencies in two aspects.
First, message passing between hosts introduces non-negligible overhead caused by message enqueue/dequeue and data copy. %

Second, the overall throughput suffers from load imbalance among hosts~\cite{netcache2017,Elmore2011ZephyrLM,Kulkarni2017RocksteadyFM}.

\emphbf{Disaggregated memory (DM) indexes (\autoref{fig:index-dm}) are inefficient.}
Shared-everything DM indexes allow all hosts to operate on a single shared index located on memory nodes using one-sided RDMA.
However, DM indexes are over-designed for low-latency and CPU-cacheable memory interfaces.
For example, clients of DM indexes usually use a CCIndex to store locations of data~\cite{sherman2021,luo2023SMART,shen2023fusee}.
With shared memory on PCC platforms, the client-side indexes prolong the lookup latency and waste memory on each client (\autoref{fig:eval-ycsb}).

\subsection{Our approach for building \ds}
We adopt a two-step design approach for building \ds.
First, we convert existing CCIndex to PCCIndex, aiming to explore how to achieve correctness with minimal effort.
Then, we optimize the converted indexes by identifying and mitigating efficiency bottlenecks.
We choose to build upon CCIndexes because they are highly efficient. Additionally, we can reduce the development efforts by leveraging the mature designs and optimizations already present in CCIndexes.

\label{sec:moti-requirements}

We detail \textbf{key requirements} of \ds as follows.

\begin{itemize}
\item
\textbf{R1: concurrent-safety (i.e., linearizability~\cite{Herlihy1990Linearizability}, \autoref{sec:conversion-r1}).}
\ds must ensure concurrent-safety, which means it behaves correctly when accessed or manipulated by concurrent threads.
This paper uses linearizability~\cite{Herlihy1990Linearizability} as the correctness condition for concurrent threads.

\item
\textbf{R2: crash-consistency (\autoref{sec:conversion-r2}).}
\ds must ensure crash consistency as its state persists in shared memory, which remains accessible after failures of processing threads/hosts~\cite{zhang2023partial,zhu2024Lupin}.
First, \ds should ensure \textbf{durable linearizability (R2.1)}, which means that operations should complete in an all-or-nothing manner to ensure that indexes remain in a consistent state upon crashes.
Second, \ds should ensure \textbf{failure-isolation (R2.2)}, meaning that the failure of one host must not affect other active hosts.

\item
\textbf{R3: efficiency and scalability (\autoref{sec:guidelines}).}
\ds should strive to be efficient and scalable.
\end{itemize}

\section{Consistent \ds with \Rules}
\label{sec:approach}

This section presents how we convert existing CCIndex to PCCIndex, while ensuring concurrent-safety (R1, \autoref{sec:conversion-r1}) and crash-consistency (R2, \autoref{sec:conversion-r2}) of PCCIndex.

\subsection{Making \ds Concurrent-safe (R1)}%
\label{sec:conversion-r1}

\subsubsection{High-level Approaches: \Rules}
\label{sec:conversion-r1-sp-rules}

To ensure linearizability of the \ds with the absence of cache coherence, we must correctly synchronize index's shared data in CPU cache.
Specifically, we must ensure that once shared data becomes visible to other hosts, no host should read stale data or miss the newly updated data.

We observe that different shared data have different requirements for update visibility.
There are two types of shared data in an index, (1) \emph{sync-data}: the data used for synchronization (e.g., the \emph{lock flag} for lock-based indexes or the \emph{version} field for optimistic lock-based indexes~\cite{Masstree2012,leis2016ARTOLC}). 
(2) \emph{protected-data}: the data protected by synchronization mechanisms (e.g., nodes).
\emph{Sync-data} must be immediately visible to others upon changes to avoid inconsistency. 
Taking a lock-based index as an example, the \emph{lock flag} must be immediately visible to prevent multiple cores from getting the lock simultaneously.
In contrast, \emph{protected-data} does not require immediate visibility, as long as the synchronization mechanisms protect the data from being accessed by other cores.
For example, data modified within a critical section only needs to be visible to other cores before the critical section ends.

\textbf{\Rules.} We propose to synchronize \emph{\underline{S}ync-data} and \emph{\underline{P}rotected-data} differently.
PCC platforms have two methods to synchronize data: (1) bypass-cache \pcas/\pload (defined in \autoref{sec:moti-pcc-model}) ensures immediate visibility of 8-byte data, and (2) cacheline control operations can control data visibility by invalidating data in CPU cache through \texttt{clflush+mfence}%
\footnote{Cacheline invalidation can be done using the standard \texttt{clflush+mfence} operation, as \texttt{clflush} will not write back stale data that has not been updated according to Intel~\cite{intel-manual} and AMD~\cite{amd-manual} specifications.
}%
or writing back data to shared memory through \texttt{clwb+mfence}.
To ensure concurrent-safety, \emph{sync-data} should be manipulated using \pcas/\pload to ensure immediate visibility.
\emph{Protected-data} can be temporarily stored in the cache, but protected-data's cacheline must be invalidated (i.e., \texttt{clflush+mfence}) before reading the data and written back (i.e., \texttt{clwb+mfence}) after writing the data to ensure visibility.

\subsubsection{Concrete Examples}
\label{sec:conversion-r1-conversion-methods}

\begin{figure}[t]
    \centering
    \begin{minipage}[t]{1\linewidth}
        \centering
        \includegraphics[width=.99\linewidth]{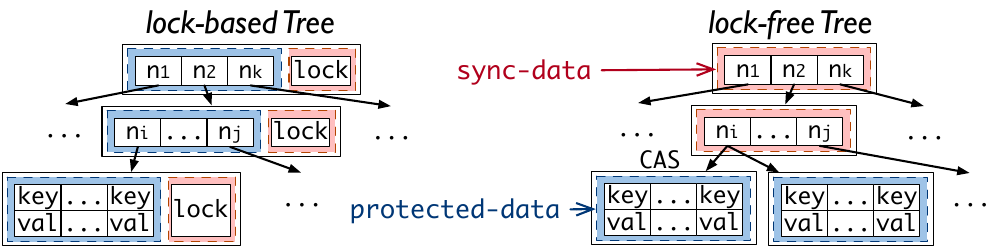}
        \vspace{3px}
    \end{minipage}
    \begin{minipage}[t]{1\linewidth}
        \begin{minipage}[t]{0.49\linewidth}
            \centering
            \includegraphics[width=1\linewidth]{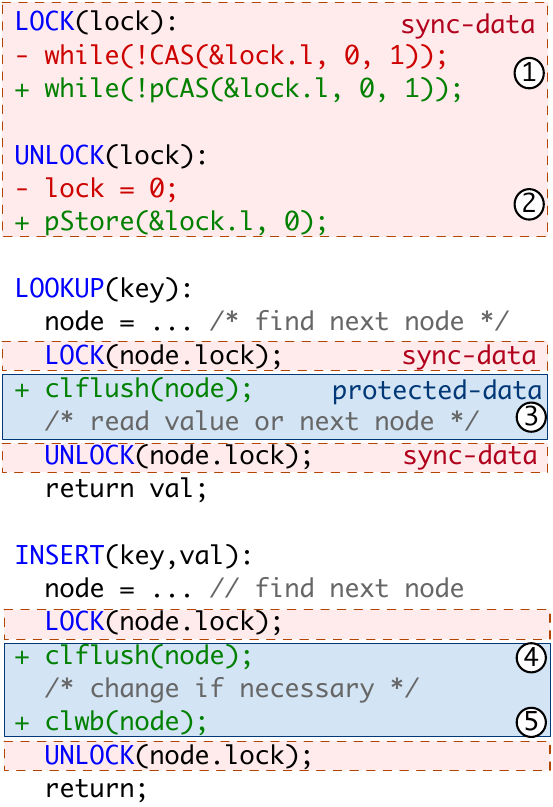}
            \vspace{-20px}
            \subcaption{Lock-based}
            \label{fig:code-lock-based}
        \end{minipage}
        \hfill
        \begin{minipage}[t]{0.49\linewidth}
            \centering
            \includegraphics[width=1\linewidth]{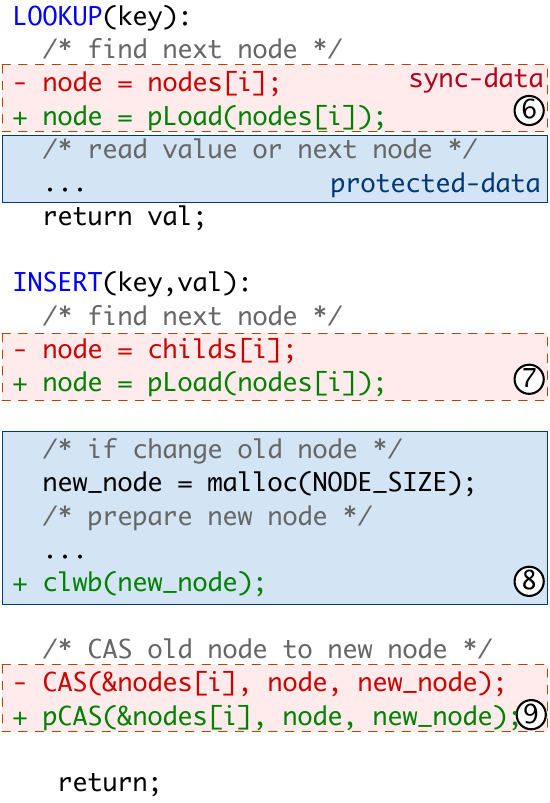}
            \vspace{-20px}
            \subcaption{Lock-free}
            \label{fig:code-lock-free}
        \end{minipage}
    \end{minipage}
    \vspace{-12px}
    \caption{\textbf{Pseudo code of lock-based and lock-free \ds.}
    Sync-data is synchronized by \pcas/\pload, and protected-data is synchronized by \texttt{clflush}/\texttt{clwb}.
        \emph{\autoref{sec:appendix-code-indexes} futher shows detailed code of a OLC-based and a lock-free \ds.}
    }
    \vspace{-10px}
\label{fig:code-consistency}
\end{figure}

\textbf{Lock-based index (\autoref{fig:code-lock-based}).}
The sync-data of lock-based index is the \emph{lock flag}, which uses \pcas/\pload for locking and unlocking (\ding{172}, \ding{173}). 
Besides, the protected-data is node data, which is protected by cacheline flush of the entire node.
Before accessing a node, we must first invalidate the node's cacheline from CPU cache with \texttt{clflush} + \texttt{mfence} to avoid reading stale data (\ding{174}, and \ding{175}). 
After updating a node, we write back the cacheline with \texttt{clwb} + \texttt{mfence} to ensure the changes in CPU cache are written back and visible to other hosts (\ding{176}).

\textbf{Lock-free index (\autoref{fig:code-lock-free}).}
Lock-free indexes perform out-of-place node updates. 
As shown in \autoref{fig:code-lock-free}'s \texttt{INSERT} pseudo code, to update a node, a new node is allocated (\ding{179}), and a \texttt{CAS} operation atomically updates the node pointer from the old node to the newly allocated node (\ding{180}). 
The sync-data of lock-free indexes is the node pointers, which are updated by \pcas (\ding{180}) and loaded by \pload (\ding{177} and \ding{178}).
With \pcas, concurrent \pload observe either the old or the new node pointer, ensuring consistency. 
The protected-data is node data.
Accesses to protected-data do not require invalidating the cacheline first, since nodes are never updated in place and each machine will not hold stale cacheline.
The newly allocated node must be written back to shared memory to ensure visibility to other hosts.
The write-back is done through \texttt{clwb}+\texttt{mfence} (\ding{179}) before performing  \pcas that updates the node pointer.

\textbf{Indexes using other concurrency control mechanisms} like OLFIT~\cite{Cha2001OLFIT}, optimistic lock coupling (OLC)~\cite{Masstree2012, leis2016ARTOLC}, ROWEX~\cite{leis2016ARTOLC} can also be converted to \ds following \rules.
For example, OLC-based indexes' sync-data is the \emph{version} field, which serves as a lock for each node.
For writer, it sets a lock-bit in the \emph{version} to acquire the lock of a node, and increments the \emph{version} when updating a node to notify readers.
For reader, it checks the \emph{version} before and after reading a node, and retries the operation if the two accessed versions do not match.
OLC-based indexes' protected-data is the node data, which is synchronized similarly to lock-based indexes, i.e., invalidating the cacheline before reading, and writing-back the cacheline after writing the data.

\subsubsection{Additional Requirements for Correctness}
\label{sec:sp-rules-additional-requirements}
There are two additional requirements for correctness:

(1) Each node must be placed in a separate cacheline.
Otherwise, two independent node write-backs could corrupt each other's data.
Since most indexes' node are larger than a cacheline (e.g., 4KB for B-tree) and aligned to a cacheline, this requirement will not introduce memory overhead.

(2) For lock-free indexes, the newly allocated node must not be pre-cached on other machines. Otherwise, stale data will be read since lock-free indexes do not invalidate cacheline before reading the data.
Ensuring data is not pre-cached includes two requirements. First, the hardware should not prefetch data to the cache.
This can be achieved by disabling hardware cache prefetching, and in the case of indexing, disabling hardware cache prefetching has little performance impact since indexes access memory in a random order.
Second, the cacheline of deleted nodes must be invalidated on all hosts before reallocating.
To implement this, we send message to each host to ask background threads to flush the cacheline of nodes to be deleted. 
After all hosts confirm the flush, we proceed with memory reallocation.
This operation is not on the critical path, and has little impact on performance.

\subsection{Making \ds Crash-consistent (R2)}%
\label{sec:conversion-r2}

\textbf{Durable Linearizability (DL).}
To ensure DL of \ds, we apply the techniques developed for persistent memory (PM) indexes~\cite{Lee2019RecipeCC}.
Existing studies~\cite{Lee2019RecipeCC,Ramanathan2021TIPSMV} have shown that many DRAM indexes can be converted to PM indexes that guarantee DL by (1) inserting \texttt{clwb} + \texttt{mfence} after each store operation, and (2) after crashes, \texttt{INSERT/LOOKUP} operations use helper mechanisms to fix inconsistency.
We ensure DL of \ds using PM index's technique, i.e., inserting \texttt{clwb} + \texttt{mfence}. 
These cacheline write-back operations both ensure concurrency safety (R1, \autoref{sec:conversion-r1}) and DL (R2.1) of \ds.

\noindent
\textbf{Failure Isolation.}
To ensure failure isolation (R2.2), \ds should prevent a failed worker thread from crashing or blocking other threads.
Lock-free indexes naturally do not crash/block other threads since all updates are conducted through an atomic \pcas, so the index is always in a consistent state.
In contrast, update operations on lock-based indexes acquire locks, which may block other threads if the locks are not properly released.
To address the blocking problem of lock-based indexes, we use \emph{recoverable lock}s plus the controller-supported host heartbeat checking inspired by Lupin~\cite{zhu2024Lupin}.
Specifically, we use a 16-bit field in the 64-bit \emph{lock flag} to record the host ID of the lock owner.
When an alive host attempts to acquire the lock, if it retries many times, and once it exceeds a \emph{timeout} (e.g., 1\,ms), it will ask the controller whether the lock's owner is still alive.
If the lock's owner is not alive, the controller will release the lock by clearing the \emph{lock flag}.
The controller uses heartbeat to check host liveness.

\section{Efficient \ds with \guideline Guidelines}%
\label{sec:guidelines}

This section analyzes the performance overhead of \ds, and presents \gl based on the observations.

\subsection{Latency Overhead}
\label{sec:sig-op-overhead}

We first break down the latency of \ds's \texttt{LOOKUP} and \texttt{INSERT} operations.
Specifically, the overhead consists of (1) cache-bypass primitives (i.e., \pload/\pstore/\pcas), which have higher latency compared to \texttt{Load}/\texttt{Store}/\texttt{CAS}, and (2) cacheline control operations (i.e., \texttt{clwb}/\texttt{clflush} + \texttt{mfence}).

\begin{table}[t]
\centering  
\caption{\textbf{Overhead of converting CCIndex to {\ds}.}
LF: lock-free, OLC: optimistic lock coupling.
In column \emph{PCC Latency}, the first value is the average latency of {\ds} and the second is the overhead introduced by the conversion.}
\vspace{-8px}
\label{tab:indexes}
{\setlength{\tabcolsep}{3pt}%
\resizebox{\linewidth}{!}{%
\begin{tabular}{@{}lllrr@{}}
    \hline
    \multicolumn{1}{c}{\multirow{2}{*}{\textbf{Index Name}}} & \multicolumn{1}{c}{\multirow{2}{*}{\textbf{Structure}}} & \multicolumn{1}{c}{\multirow{2}{*}{\textbf{Type}}}  & \multicolumn{2}{c}{\underline{\textbf{PCC Latency (Overhead)}}} \\
    \multicolumn{1}{c}{}                                  & \multicolumn{1}{c}{}                             & \multicolumn{1}{c}{} & \multicolumn{1}{r}{\textbf{lookup ($\mu$s)}} & \multicolumn{1}{r}{\textbf{insert ($\mu$s)}} \\ \hline
    \rowcolor{gray!30}
    CLHT~\cite{David2015ASCY} & Hash Table & LF+Lock & 1.9 (1.9) & 2.2 (2.2) \\
    \rowcolor{gray!30}
    CLevel-Hash~\cite{chen2020CLevel} & Hash Table & LF & 9.9 (9.8) & 4.4 (4.1) \\
    BTreeOLC~\cite{leis2016ARTOLC}& B+ Tree   & OLC & 9.3 (9.1) & 10.4 (10.2) \\
    \rowcolor{gray!30}
    BwTree~\cite{Levandoski2013BwTree}  & B+ Tree   & LF & 2.5 (2.1) & 4.6 (3.9) \\
    HOT~\cite{Binna2018HOT} & Trie      & OLC & 10.3 (10.1) & 23.5 (23.2) \\
    ARTOLC~\cite{Leis2013ART,leis2016ARTOLC}  & Radix Tree & OLC & 7.3 (7.2) & 46.5 (45.9) \\
    Masstree~\cite{Masstree2012}  & Hybrid Tree & OLC & 38.3 (38.0) & 39.7 (39.2) \\
    \hline
\end{tabular}
}
}%
\vspace{-5px}
\end{table}

\textbf{Test.}
\autoref{tab:indexes} lists the indexes we converted and tested, together with the average latency of \ds, and the latency overhead of \ds compared with CCIndexes.
We run each \ds with one thread and one million 8-byte/8-byte key/value pairs on CXL memory.
According to the results, OLC-based indexes suffer from significant latency increases (7.2 $\mu$s to 39.2 $\mu$s), while lock-free indexes exhibit the least additional overhead (1.9 $\mu$s to 9.8 $\mu$s).
Besides, the absolute latency of lock-free \ds is also lower than CCIndexes.

\textbf{Analysis.}
We identify the primary benefit of lock-free indexes as their use of out-of-place updates.
First, out-of-place updates eliminate the need for invalidating \emph{protected-data}'s cacheline before reading.
This is because out-of-place updates do not modify existing objects, so cached copies of objects are always up-to-date.
In contrast, OLC-based indexes perform in-place modifications, thereby necessitating cacheline invalidation before accessing \emph{protected-data}.
Second, out-of-place updates allow lock-free indexes to access and update \emph{sync-data} (e.g., node pointers) with only a single \pload and a single \pcas operation, respectively. 
In contrast, OLC-based indexes use in-place updates, and acquiring/releasing read-lock requires at least two \pload operations on the \texttt{version} field, while acquiring/releasing write-lock requires one \pcas and one \pstore operation to modify the \texttt{version} field.

\myboxA{
\underline{\textbf{Observation \#1}}: Out-of-place updates are better than in-place updates on PCC platforms, as out-of-place updates result in fewer cache-bypass operations and cacheline flushes.
}

\myboxB{
    \textbf{G1}: We suggest converting indexes using \textbf{\emph{\underline{O}ut-of-place updates}} to minimize costly cacheline control operations.
}

\subsection{Scalability Bottleneck}
\label{sec:scalability-overhead}

As we increase the concurrency, we found that {\ds} shows poor scalability (see \emph{PCC-BwTree} in \autoref{fig:eval-ycsb} for specific data).
A detailed breakdown reveals that the primary performance bottleneck stems from concurrent \pload of heavily accessed shared variables. 
For instance, each operation in BwTree requires a \pload of the root node pointer, which accounts for more than 80\% of the total operation latency.

\begin{figure}[t]
    \centering
    \includegraphics[width=1\linewidth]{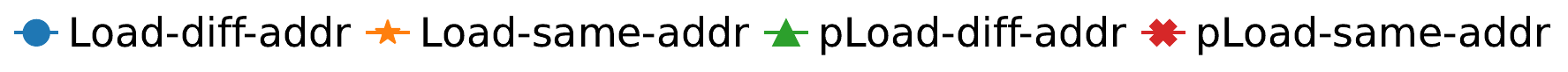}
    \vspace{-2px}
\begin{minipage}[t]{\linewidth}
    \centering
    \begin{minipage}[t]{0.34\linewidth}
        \centering
        \includegraphics[width=\linewidth]{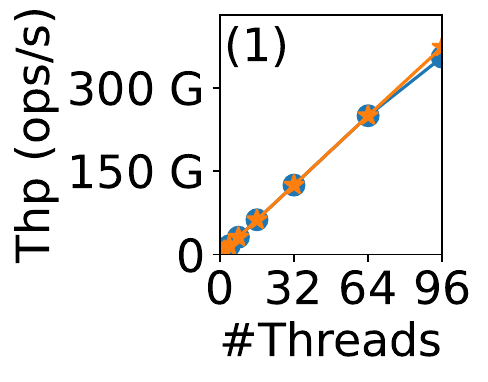}
    \end{minipage}
    \begin{minipage}[t]{0.32\linewidth}
        \centering
        \includegraphics[width=\linewidth]{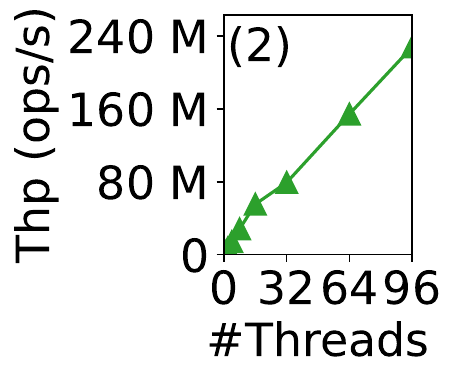}
    \end{minipage}
    \begin{minipage}[t]{0.3\linewidth}
        \centering
        \includegraphics[width=\linewidth]{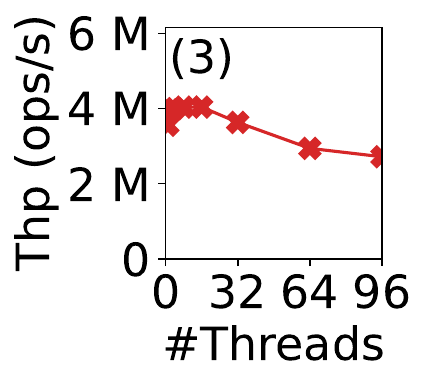}
    \end{minipage}
    \vspace{-8px}
    \subcaption{\textbf{Throughput.}
    (1) shows Load-diff-addr and Load-same-addr; (2) shows pLoad-diff-addr; (3) shows pLoad-same-addr.}
    \label{fig:moti-parallel-rw-thp}
\end{minipage}
\begin{minipage}[t]{\linewidth}
\centering
    \includegraphics[width=.96\linewidth]{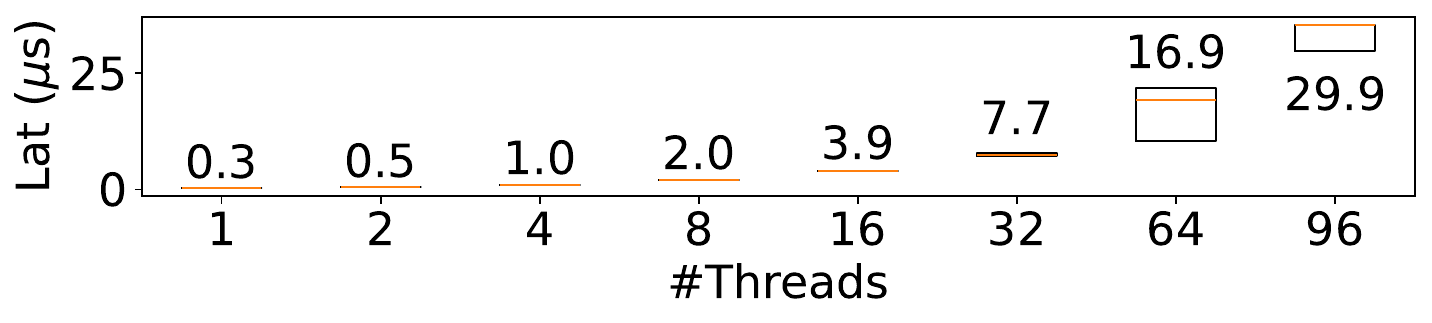}
    \vspace{-8px}
    \subcaption{\textbf{Latency of pLoad-same-addr.} The text shows P50 latency.}
    \label{fig:moti-parallel-rw-lat}
\end{minipage}
\vspace{-8px}
\caption{\textbf{Concurrent \texttt{\pload}/\texttt{Load} of same/different physical addresses.}
Concurrent \pload of the same physical address (pLoad-same-addr) shows poor scalability.
}
\label{fig:moti-parallel-rw}
\vspace{-5px}
\end{figure}

\textbf{Test.} We conduct tests (\autoref{fig:moti-parallel-rw}) to unravel why \pload bottleneck the scalability of \ds.
We test the throughput/latency of concurrent \pload of the same address (\emph{pLoad-same-addr}), and for comparison, we test the throughput/latency of cached \texttt{Load} of same or different physical addresses (\emph{Load-same-addr} and \emph{Load-diff-addr}), as well as \pload of different physical addresses (\emph{pload-diff-addr}).
As shown in \autoref{fig:moti-parallel-rw-thp}, all cases except \emph{pLoad-same-addr} scale nearly linearly with increasing concurrency. 
In contrast, the scalability of \emph{pLoad-same-addr} is significantly limited.
\autoref{fig:moti-parallel-rw-lat} further shows that as the number of threads increases, the P50 latency of \emph{pLoad-same-addr} rises sharply from 0.3\,$\mu$s (1 thread) to 29.9\,$\mu$s (96 threads).
By contrast, the P50 latency for the other three cases remains stable: for \emph{Load-same-addr} and \emph{Load-diff-addr}, it is 0.3\,$\mu$s from 1 to 96 threads; for \emph{pLoad-diff-addr}, it is 0.3\,$\mu$s to 0.4\,$\mu$s from 1 to 96 threads.

\textbf{Analysis.} 
We found that the fundamental cause of poor scalability is the \emph{strict memory ordering requirement} enforced on existing platforms for preserving memory semantics (according to Intel manual~\cite{intel-manual}). 
Specifically, memory operations must adhere to a strict execution order, prohibiting even the reordering of read operations. 
Under this constraint, the hardware memory controller must execute all read operations strictly in the order they are issued~\cite{midhul2024understanding}, requiring each to await the completion of preceding requests. 
In contrast, cache coherence protocols allow cached \texttt{Load}s to bypass memory controller serialization, since it helps determine whether a memory location has been modified.

\myboxA{
    \underline{\textbf{Observation \#2}}: On PCC platforms, concurrent \pload of the same physical address (e.g., index's frequently-accessed variables) are serialized due to strict memory ordering, which creates a scalability bottleneck.
}

\textbf{Solution.}
Indexes on PCC platforms should avoid using concurrent \pload of the same physical address.
Most existing CCIndexes rely on frequently-accessed shared variables (e.g., the pointer to the root node for tree-structure indexes, global information like bucket location for hash tables). 
We should identify the frequently-accessed variables of indexes, and avoid \pload of these variables.
One simple and feasible way is to replicate them, so that each thread can \pload different replicas to avoid the performance penalty of concurrent \pload same physical address.

\myboxB{
    \underline{\textbf{G2}}: Utilize \textbf{\emph{\underline{R}eplicated shared variables}} to avoid concurrent \pload of the same physical address.
}

\subsection{Inefficiency under Read-heavy \& Skewed Workloads}
\label{sec:real-trace-overhead}

The performance gap between \ds and its CCIndex counterpart becomes more significant under read-heavy and skewed workloads, and workloads with such features are common in real-world production systems~\cite{berk2012workload-analysis,nishtala2013scaling,Yang2020TwitterTrace}.

\begin{figure}[t]
    \centering
    \includegraphics[width=.98\linewidth]{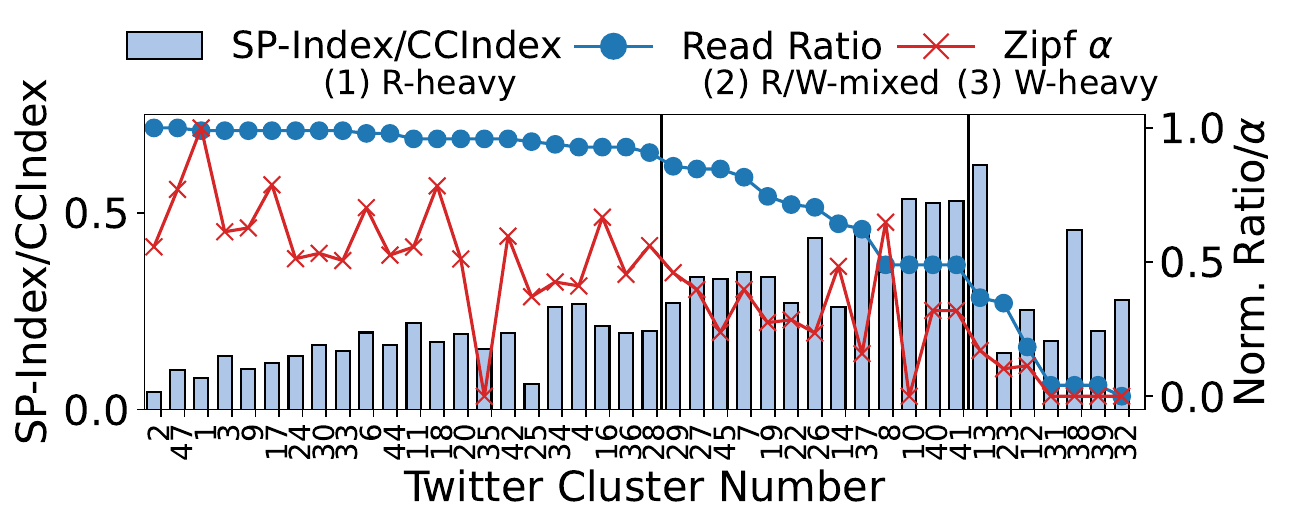}
    \vspace{-12px}
    \caption{\textbf{Performance gap between PCCIndex and CCIndex in Twitter traces~\cite{Yang2020TwitterTrace}.} 
    The performance gap widens (indicated by higher CC-Thp/PCC-Thp) as traces become more read-heavy (higher read ratio) and more skewed (higher zipf $\alpha$).
    zipf is normalized to 3, traces are sorted by read ratio.
    }
    \vspace{-5px}
    \label{fig:obs-read-heavy}
\end{figure}

\textbf{Test.}
To evaluate \ds under read-heavy and skewed workloads, we chose one of our converted \ds (i.e., a converted BwTree~\cite{wang2018openbwtree}) and evaluated it on 42 real-world Twitter traces~\cite{Yang2020TwitterTrace}.
Performance gap is indicated by \emph{CC-Thp/PCC-Thp} in \autoref{fig:obs-read-heavy}, which denotes the throughput ratio of CCIndex to PCCIndex.
Each trace's read ratio (\emph{read ratio}) and skewedness (\emph{norm. zipf $\alpha$}) are also marked in the figure.
The result shows that the performance gap widens as traces become more read-heavy and more skewed.

\textbf{Analysis.}
The widening performance gap primarily stems from the fact that read-heavy and skewed workloads benefit a lot from the extremely low latency of the CPU cache (<10\,ns), , whereas \ds must maintain consistency by performing {\pload} operations (300--500\,ns), even when the data is cached and unchanged.
Specifically, on cache-coherent platforms, frequently accessed and immutable keys are effectively retained in the CPU cache, resulting in a low cache miss rate.
For example, running CCIndex (BwTree in \autoref{fig:obs-read-heavy}) on Twitter trace \#1 (read ratio=99\%, zipf $\alpha$=2.67) exhibits a miss rate of only 0.2\% and achieves lookup latency as low as \textasciitilde 100\,ns, since most data accesses are cache hits.
In contrast, operations on \ds require {\pload} even for cached and unchanged data. 
For instance, a PCCIndex (BwTree) must perform {\pload} on the pointer of each traversed node, resulting in a latency of \textasciitilde 2500\,ns since it uses 6 \pload operations.

\myboxA{
    \underline{\textbf{Observation \#3}}: The performance gap between PCCIndex and CCIndex widens as traces become more read-heavy and skewed, since \ds must perform {\pload} even for cached and unchanged data.
}

\textbf{Solution.}
Under read-heavy and skewed workloads, \ds can enhance performance by speculatively \texttt{Load}ing cached (but potentially stale) data from local caches, rather than immediately {\pload} shared data. 
In this approach, \ds's operations first follow a fast path that \texttt{Load}s data from the local cache; if the data is determined to be stale at any point along the access path, the system falls back to a slower path that performs {\pload} to ensure data freshness. 

The \emph{performance benefits} of speculative reading arise from two key optimizations.
First, \emph{{\pload} operations are reduced.} Since we only need to detect and fix stale data at some point, multiple costly {\pload} can be replaced with a final {\pload} to detect staleness. 
    Note that speculative reading is particularly beneficial for read-heavy and skewed workloads, as in other scenarios the cost of retries may outweigh its advantages. 
Second, CPU caches on the same host are coherent, which allows worker threads to reuse cached data, further cutting down the number of {\pload} for updating the cached data.

\myboxB{
    \underline{\textbf{G3}}: Leverage \textbf{\emph{Sp\underline{E}culative Reading}} instead of \pload every data to utilize the fast CPU cache.
}

\subsection{{\guideline} Guidelines and Correctness}
\label{sec:guidelines-guidelines}

To conclude, we propose \guideline guidelines: \underline{O}ut-of-place update (G1), \underline{R}eplicated shared variable (G2), Sp\underline{E}culative Reading (G3).
We first discuss how to ensure correctness of G2/G3.

\subsubsection{Ensure Correctness of G2 (Replicated Variables)}
\label{sec:guidelines-fix-inconsistency-g2}

\begin{figure}[h]
    \centering
    \begin{minipage}[t]{\linewidth}
        \vspace{-8px}
        \includegraphics[width=.85\linewidth]{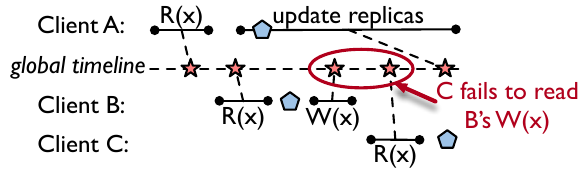}
        \vspace{-5px}
        \subcaption{Inconsistency: B's W(x) and C's R(x) are not linearizable.}
        \label{fig:replication-linearizability-inconsistency}
    \end{minipage}
    \begin{minipage}[t]{\linewidth}
        \includegraphics[width=.88\linewidth]{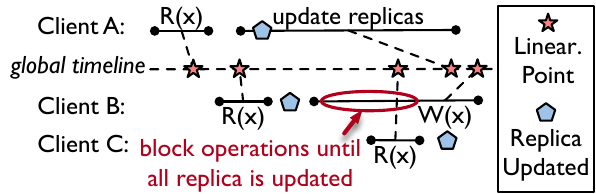}
        \vspace{-5px}
        \subcaption{Fix: block operations starting from newer replicas during update.}
        \label{fig:replication-linearizability-fix}
    \end{minipage}
    \vspace{-8px}
    \caption{\textbf{Ensure correctness of G2.}}
    \label{fig:replication-linearizability}
    \vspace{-5px}
\end{figure}

Replicating shared variables following G2 can lead to inconsistencies since replicas cannot be updated atomically.
As illustrated in \autoref{fig:replication-linearizability-inconsistency}, operations that begin with stale local replicas (e.g., Client C's Read(x)) may miss updates performed after replica synchronization (e.g., Client B's Write(x)), thereby violating linearizability.

To fix the inconsistency, we proposed to block all operations that start with newer replicas until all replicas are updated.
The blocking is feasible to implement: one simple way is to use the last bit of the replica as a lock to block operations starting from newer replicas during update.
Detailed implementation is stated in \autoref{sec:orohash-replicate-context} and \autoref{sec:orbtree-replicate-gcid}.
Besides, the overhead of blocking is acceptable, since G2 only replicates shared variables, and the updates of these variables are really rare.
For example, the tree's root node is changed only during structural modification, the hash table's global context is changed only during rehashing (inserting 100 million keys introduces only 8 times resize for our tested hash table).

\subsubsection{Ensure Correctness of G3 (Speculative Reading)}
\label{sec:guidelines-fix-inconsistency-g3}

G3 allows readers to read stale data, which causes inconsistency since stale data is not the latest.
\emph{Correctness} is guaranteed by: (1) the system must always be able to detect if stale data has been read, and (2) no incorrect behavior (i.e., access freed memory) occurs during read operations processing with stale data.
Detailed implementation is described in \autoref{sec:orobtree-opt-read}.

\section{Case Studies}%
\label{sec:design}

We apply \rules and {\gl}s to: (1) a hash table, i.e., CLevel-Hash~\cite{chen2020CLevel}, (2) a B+Tree, i.e., BwTree~\cite{wang2018openbwtree}, and (3) a decentralized garbage collection~\cite{tu2013silo,wang2018openbwtree}.
(1) and (2) use out-of-place updates, making them adhere to \textbf{G1}.

\begin{figure}[t]
    \centering
    \begin{minipage}[t]{0.51\linewidth}
        \includegraphics[width=1\linewidth]{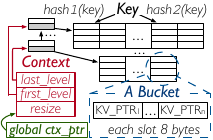}
        \vspace{-15px}
        \subcaption{Original CLevel-Hash}
        \label{fig:tech-hash-basic}
    \end{minipage}
    \begin{minipage}[t]{0.43\linewidth}
        \includegraphics[width=1\linewidth]{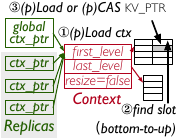}
        \vspace{-15px}
        \subcaption{Operations w/o \texttt{REHASH}}
        \label{fig:tech-hash-read}
    \end{minipage}
        \begin{minipage}[t]{1\linewidth}
        \includegraphics[width=1\linewidth]{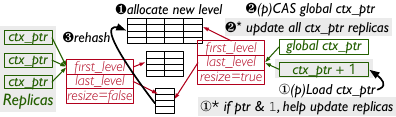}
        \vspace{-18px}
        \subcaption{Operations during \texttt{REHASH}}
        \label{fig:tech-hash-rehash}
    \end{minipage}
    \vspace{-10px}
    \caption{\textbf{Apply \rules and G2 to ClevelHash.}
    Following G2, we replicate the \emph{global ctx\_ptr} as per-thread \emph{ctx\_ptr} replicas. 
    We modify steps \ding{172} and \ding{183} to ensure consistency.
    }
\label{fig:tech-orohashtable}
\vspace{-5px}
\end{figure}

\subsection{Case Study \#1: ClevelHash}%

\textbf{How original ClevelHash works.}
CLevel-Hash~\cite{chen2020CLevel} is a lock-free hash table.
As \autoref{fig:tech-hash-basic} shows, CLevel-Hash has several levels, each containing a set of buckets.
A \emph{global ctx\_ptr} variable points to the \emph{global context}, which tracks the first and last levels, and whether the hash table is resizing.
To find a key (as shown in \autoref{fig:tech-hash-read}), operations \texttt{Load} the global context (\ding{172}), traverse from the last level to the first level to find the target slot (\ding{173}), and read or update the key/value pointer (i.e., $KV\_PTR_{i}$) in the slot with \texttt{Load} and \texttt{CAS} (\ding{174}).
To extend the hash table (as shown in \autoref{fig:tech-hash-rehash}), a new level is allocated (\ding{182}) and the global context is updated out-of-place, i.e., \texttt{CAS} \emph{global ctx\_ptr} to a newly allocated context with \emph{first\_level} pointing to the new level and \emph{resize=true} (\ding{183}).
The background \texttt{REHASH} thread then rehashes entries from the last level to upper levels (\ding{184}).
After all entries are rehashed, the \emph{global ctx\_ptr} is updated again, pointing to a context with the new last layer and \emph{resize=false}.

\subsubsection{Apply \Rules}
Following \rules,
we identify the \emph{global ctx\_ptr} and key/value pointers ($KV\_PTR_{i}$) in bucket slots as sync-data.
We manipulate these pointers using \pcas/\pload (\ding{172},\ding{174} \autoref{fig:tech-hash-read}; \ding{183} in \autoref{fig:tech-hash-rehash}).
The global context pointed by \emph{global ctx\_ptr} and key/value data pointed by $KV\_PTR_{i}$ are protected-data.
As they are updated out-of-place, we flush them once after creation, and directly access them afterwards.

\subsubsection{Apply \guideline-G2: Replicated Context Pointer}%
\label{sec:orohash-replicate-context}

\textbf{Replicate \emph{global ctx\_ptr}.}
After conversion with \rules, concurrent \pload of \emph{global ctx\_ptr} (\ding{172} in \autoref{fig:tech-hash-read}) becomes a scalability bottleneck.
To mitigate this, we apply \textbf{G2} by replicating the \emph{global ctx\_ptr} for each worker thread (referred to as \emph{ctx\_ptr} in \autoref{fig:tech-hash-read}), allowing each worker thread to \texttt{pLoad} a different \emph{ctx\_ptr}.
These replicated \emph{ctx\_ptr}s are updated to the new \emph{global ctx\_ptr} after the \emph{global ctx\_ptr} itself is updated (\ding{183}* in \autoref{fig:tech-hash-rehash}).
The \emph{ctx\_ptr} replicas are also placed on shared memory, enabling every host to access them with \pload (\ding{172}) and update them in \ding{172}*.

\noindent\textbf{Ensure consistency of \emph{ctx\_ptr} replicas.}
The \emph{ctx\_ptr} replicas may become inconsistent since they cannot be updated atomically with the \emph{global ctx\_ptr}.
To fix the inconsistency, we block all operations that start with newer \emph{ctx\_ptr} until all replicas are updated (this method is also discussed in \autoref{sec:guidelines-fix-inconsistency-g2}).
Specifically, we use the \emph{ctx\_ptr's last bit} as a lock\footnote{The last bit is 0 by default since all replica pointers are 8-byte aligned}: setting it to \texttt{1} indicates there exists an in-flight replica update, thus, any operation detecting this will help update all replicas before proceeding with its own logic ((\ding{172}* in \autoref{fig:tech-hash-read})).
The help mechanism works as follows: the operation compares each replica with the \emph{global ctx\_ptr} and updates any outdated replicas to the latest \emph{global ctx\_ptr}; once all replicas are updated, the operation clears the lock bit and continues.
If the \emph{global ctx\_ptr} is modified during this process, the help mechanism detects the change when comparing replicas and initiates another round of updates. 
This process will eventually complete, provided that updates to the \emph{global ctx\_ptr} eventually stop.

\textbf{Durable Linearizability (DL).} Furthermore, replicated \emph{ctx\_ptr} does not compromise the DL of ClevelHash.
In ClevelHash without replicas, all changes on shared memory ensure DL.
When replication is introduced, partly updated replicas can cause temporary inconsistencies, but all replicas can eventually update to the \emph{global ctx\_ptr}, which consistently represents the most recent and correct state. 

\subsection{Case Study \#2: BwTree}%

\begin{figure}[t]
    \centering
    \includegraphics[width=.95\linewidth]{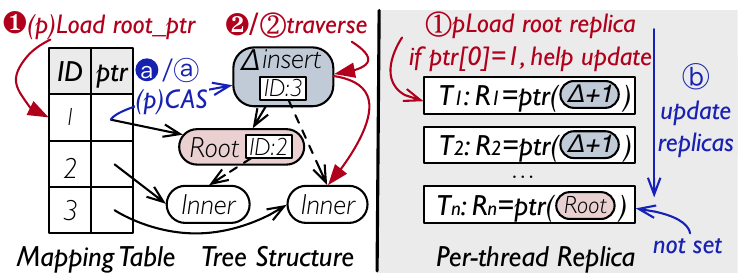}
    \vspace{-5px}
    \caption{\textbf{Apply G2 to BwTree.}
    Concurrent \pload root pointer is a scalability bottleneck (\redding{182}).
    Right part apply G2 to replicate the root pointer to each worker thread (\redding{172}).
    }
    \label{fig:bwtree-replicate}
    \vspace{-5px}
\end{figure}

\textbf{How original BwTree works.}
BwTree~\cite{Levandoski2013BwTree} is a lock-free B+Tree.
As shown in \autoref{fig:bwtree-replicate}, BwTree uses a mapping table to translate a node's \emph{ID} to its corresponding \emph{Pointer} (e.g., ID 1 to root node pointer).
With the mapping table containing all pointers in the tree, access to a node is conducted by \texttt{Load} the pointer from the mapping table (\redding{182}), and updates to the BwTree are performed by \texttt{CAS} of the mapping table entry (\bluefilledcircled{a}).
All updates in BwTree are conducted out-of-place, i.e., updates to nodes are conducted by adding a $\Delta$ node to indicate the changes.
$\Delta$ can represent modification including \emph{insert}, \emph{delete} of entries, and \emph{split}, \emph{remove}, \emph{merge} of the node.
Multiple updates to the same node form a chain of $\Delta$ records.

\subsubsection{Apply \Rules}
Following the \rules, pointers in the mapping table are sync-data, which we use \pcas (\bluefilledcircled{a}) and \pload (\redding{182}) to operate; node data is protected-data, which is written-back once after creation, and directly accessed afterwards.

\subsubsection{Apply \guideline-G2: Replicate Root Node Pointer}%
\label{sec:orobtree-replicate-root}

The performance bottleneck of BwTree is the concurrent \pload of the root node pointer (\emph{root\_ptr}) in the mapping table.
We follow \emphbf{G2} to replicate \emph{root\_ptr} on each worker thread ($R_1$ to $R_n$ in \autoref{fig:bwtree-replicate}), so that each worker thread \pload its local replica to access the BwTree root (\redding{172}).
To update the root node, a thread uses \pcas to update the \emph{root\_ptr} in the mapping table to the new root (\darkblue{\circled{a}}), and then conduct replica update by setting all replicas to the latest \emph{root\_ptr} (\darkblue{\circled{b}}).

We fix inconsistency using the same method like clevelhash (\autoref{sec:orohash-replicate-context}). i.e., set the last bit of \emph{root\_ptr} replica to \texttt{1} to indicate that replication is in-flight.

\subsubsection{Apply \guideline-G3: Speculative Reading Inner Node}%
\label{sec:orobtree-opt-read}

\begin{figure}[t]
    \centering
    \includegraphics[width=0.98\linewidth]{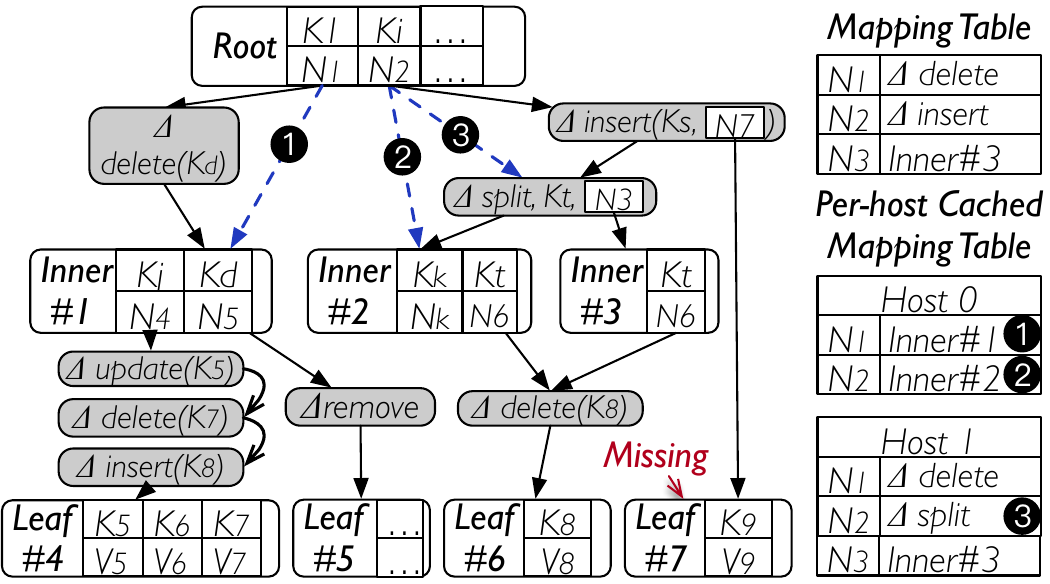}
    \vspace{-7px}
    \caption{\textbf{Apply \guideline-G3 to BwTree by speculative reading inner-node pointers.} 
    Inner-node pointers are cached in \emph{per-host cached mapping table}s, which can be stale (\ding{182}, \ding{183}, \ding{184} show three cached stale inner-node pointers).
    }
    \label{fig:tech-opt-read}
    \vspace{-5px}
\end{figure}

Since \pload of each node pointer we traverse is expensive, we allow \texttt{LOOKUP} operations to speculatively read the inner-node pointer following \emphbf{G3} (\texttt{INSERT} cannot use this optimization).
Specifically, we cache the inner-node pointer in \emph{per-host cached mapping table}.
In the fast path, \texttt{LOOKUP} traverses the tree with \texttt{Load} of inner-node pointer from the cached mapping table, and \pload leaf-node pointer in the global mapping table.
If the tree remains unchanged, readers following the fast path will get the correct value as all caches are up-to-date.
However, when updates have occurred, \texttt{LOOKUP} might read stale data, indicated by \ding{182}, \ding{183} and \ding{184} in \autoref{fig:tech-opt-read}.
Once the lookup key is not found in the fast path, \texttt{LOOKUP} falls back to \emph{the slow path}, i.e., traversing the tree by \pload pointer in the mapping table and updates pointers in \emph{cached mapping table}.

\textbf{Ensure correctness.}
As discussed in \autoref{sec:guidelines-fix-inconsistency-g3}, to ensure correctness, we must (1) always be able to detect staleness and (2) reading stale inner-node pointers should not cause crashes (e.g., access an inner-node that has been freed).

To achieve (1), retrying upon a key miss is sufficient to detect staleness because: in BwTree, inner-nodes only serve as a routing index to the leaf-nodes, and all key/value modifications are contained within the leaf node plus its associated $\Delta$ records.
\autoref{fig:tech-opt-read} shows the examples: a lookup for key=$K_d$ using $N_{1}$=Inner\#1 (\ding{182}) can still detect that Leaf\#5 has been removed, and a lookup for key=$K_8$ using $N_{2}$=Inner\#2 (\ding{183}) can detect that $K_8$ is deleted.
The only form of staleness occurs when certain leaf nodes are missing; for instance, a lookup for key=$K_9$ using $N_{3}$=$\Delta$split (\ding{184}) will miss leaf-node \#7 and, consequently, miss key $K_9$.
To address this, we enforce a slow-path retry if \texttt{LOOKUP} operation cannot find the key.

To achieve (2), the out-of-place update feature of BwTree ensures that reading a stale node remains consistent.
Furthermore, it is essential to invalidate all cached data before freeing the memory of an inner node, thereby preventing any access to the deallocated node.
This is achieved by sending messages to all hosts, asking them to set cached pointers to the node to \emph{NULL} prior to memory deallocation. This process is performed together with cache invalidation of the node (the requirement of cache invalidation is discussed in \autoref{sec:sp-rules-additional-requirements} (2)).

\emphbf{Durable Linearizability (DL).}
Additionally, cached tables do not affect DL of BwTree, as cached data can always catch up with the consistent data in shared memory.

\subsection{Case Study \#3: Garbage Collection Mechanism}%
\label{sec:orbtree-replicate-gcid}

We also port BwTree's decentralized garbage collection (DGC) method~\cite{wang2018openbwtree,tu2013silo} to correctly run on PCC platforms, and optimize it following \emphbf{G2}.
This garbage collection method is generic and can be applied to other indexes as well.

\begin{figure}[t]
    \centering
        \centering
        \includegraphics[width=1\linewidth]{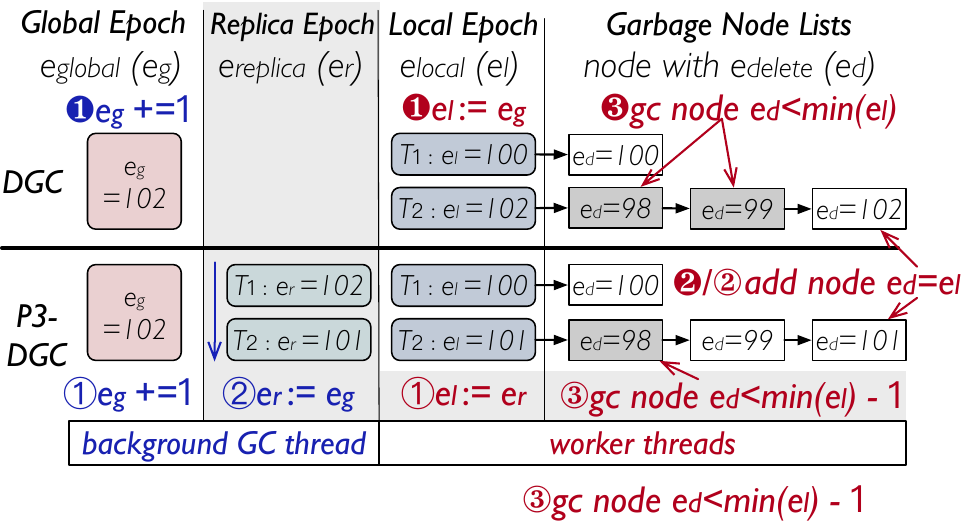}
    \vspace{-18px}
    \caption{\textbf{Apply \rules and G2 to DGC.}
    In converted SP-DGC, parallel \pload of global epoch (\darkred{\ding{182}}) will cause contention; \guideline-DGC replicates global epoch to each worker thread as $e_{r}$ (\darkred{\ding{172}}).
    Modifications are in gray box (\darkblue{\ding{173}},\darkred{\ding{172}},\darkred{\ding{174}}).
    }
    \label{fig:tech-gc}
    \vspace{-5px}
\end{figure}

\noindent\textbf{How original DGC works.}
As shown in \autoref{fig:tech-gc}, the original DGC maintains a global epoch number ($e_{g}$), while each worker thread keeps a local epoch number ($e_{l}$) and a linked list of garbage nodes that the thread has marked for deletion. 
At the beginning of each operation, a thread copies the current $e_{g}$ to its $e_{l}$ (\darkred{\ding{182}}).
When the thread deletes a node, it appends the pointer to the deleted object, tagged with $e_{d}$ equals to the latest $e_{g}$, to its list (\darkred{\ding{183}}).
After completing its operation, the thread updates $e_{l}$ to the latest $e_{g}$ again and initiates garbage collection (\darkred{\ding{184}}).
The thread retrieves the $e_{l}$ from all other threads and reclaims objects in its list that are tagged with an epoch number less than $min(e_{l})$ (\darkred{\ding{184}}).
This is safe because no threads are in the epoch that can access these objects.
To ensure progress, a background thread ($T_{gc}$) increments $e_{g}$ periodically (\darkblue{\ding{182}}).

\subsubsection{Apply \Rules}
Following the \rules, we use \pstore/\pload to update/access $e_{g}$.
$e_{l}$ should also be placed on shared memory to allow access by other threads (\redding{184} requires access to all $e_{l}$).
Other data are not shared and can be placed in local memory.

\subsubsection{Apply \guideline-G2: Replicated Global Epoch ($e_{g}$)}%
\label{sec:orobtree-replicate-gcid}

\noindent\textbf{Replicate $e_{g}$.}
\pload of the global epoch number ($e_{g}$) is a bottleneck since every operation of index must first execute \redding{182} to get the latest $e_{g}$.
We can optimize this by replicating $e_{g}$ across all GC worker threads ($e_r$) (\autoref{fig:tech-gc}), so that each GC worker thread can \pload its own $e_{r}$ (\redding{172}).
The update of replicas is conducted by the background $T_{gc}$, which increments the $e_{g}$ (\darkblue{\ding{172}}) and updates all $e_{r}$ each time (\darkblue{\ding{173}}).

\noindent\textbf{Ensure correctness.}
Replication introduces inconsistency when a thread adds a garbage node with a smaller $e_{r}$ than the latest $e_{g}$ (detailed in \autoref{sec:appendix-G2-gc-blocking}).
We fix the inconsistency of $e_{r}$ replicas by requiring that garbage nodes persist for an additional epoch before deletion.
Specifically, we alter the removal condition from ``<$min(e_{l})$'' to ``< $min(e_{l})-1$'' (\darkred{\ding{174}}).

\section{Evaluation}%
\label{sec:eval}

In this section, we evaluate \syshash, \sysbwtree (which uses \guideline-DGC), with micro-benchmarks and real-world applications to answer the following questions:
\begin{itemize}[leftmargin=1em]
\item How do \syshash and \sysbwtree perform and scale on various workloads? (\autoref{sec:eval-performance})
\item How does each technique improve the performance? (\autoref{sec:eval-factor-analysis})
\item How do real applications benefit from \ds? (\autoref{sec:eval-real-app})
\end{itemize}

\subsection{Environment Setup}
\label{sec:eval-setup}

\textbf{Testbed.}
All experiments are evaluated on a machine with 4 Intel\textsuperscript{\textregistered} Xeon\textsuperscript{\textregistered} Gold 6418H CPUs (SPR) with 4$\times$24 cores, 192/240\,MiB L2/L3 cache, and 2$\times$64\,GiB CXL 1.0 memory.
CPU frequency is fixed at 4.0\,GHz with hyper-threading enabled, Intel\textsuperscript{\textregistered} Turbo Boost disabled, and hardware cache prefetching disabled.
We use cores on NUMA node 1--3 to access CXL memory on NUMA node 0.

\noindent
\textbf{Workloads.}
We use two workloads: (1) YCSB~\cite{YCSB2010}: we select four workloads, including A (R: 50\%, W: 50\%), B (R: 95\%, W: 5\%), C (R: 100\%) and Load (W: 100\%).
We set zipfian $\alpha=0.99$ and use 100 million key value pairs (8-byte key and 8-byte value); (2) real-world Twitter~\cite{Yang2020TwitterTrace} traces: we conduct each record in the traces, using the recorded value size and operation type (lookup or insert).

\noindent
\textbf{Comparisons.}
We compare the following configurations:
\begin{itemize}[leftmargin=1em]
    \item \emph{\syshash and \sysbwtree}: CLevelHash~\cite{chen2020CLevel} and BwTree~\cite{wang2018openbwtree} with all optimizations applied.
    \item \emph{CC-CLevelHash/BwTree (ideal)}: original CLevelHash/BwTree on cache-coherence platforms with DRAM. We remove the PM-related operations in CLevelHash.
    \item \emph{MQ-CLevelHash/BwTree}: CLevelHash/BwTree on a client/server architecture which uses pass-by-reference RPC~\cite{zhang2023partial,ma2024hydrarpc}.
    We implement multi-producer single-consumer queue based on methods proposed in HydraRPC~\cite{ma2024hydrarpc}.
    We use 48 clients to send requests to 144 servers, server executes requests and returns results to clients.
    \item \emph{SP-CLevelHash/BwTree}: a converted CLevelHash/BwTree without any optimizations applied.
    \item \emph{Sherman-CXL}: We port Sherman~\cite{sherman2021}, the state-of-the-art B+Tree for RDMA-based DM architecture, to CXL platform by replacing \texttt{RDMA\_CAS} with \pcas and replacing \texttt{RDMA\_READ/WRITE} with memory copy.
\end{itemize}

\begin{figure}
    \begin{minipage}[t]{0.51\linewidth}
    \centering
    \vspace{-68px}
    \resizebox{\columnwidth}{!}{%
    \begin{tabular}{l@{\hspace{0.5em}}c@{\hspace{0.5em}}c}
    \hline
    \textbf{Memory} & \textbf{Lat.(ns)} & \textbf{B.W.(GB/s)} \\ \hline
    DRAM-L & 107 & 52 \\
    DRAM-R & 160 & 30 \\
    CXL-L & 241 & 2 \\
    CXL-R & 383 & 0.28 \\ \hline
    \end{tabular}
    }
    \subcaption{\textbf{Performance of DRAM/CXL}}
    \label{tab:eval-basic-perf}
    \end{minipage}
    \begin{minipage}[t]{0.48\linewidth}
        \centering
        \includegraphics[width=1\linewidth]{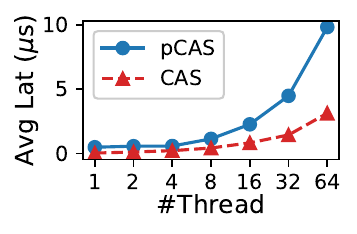}
        \vspace{-24px}
        \subcaption{\textbf{\pcas Scalability}}
    \end{minipage}
    \vspace{-16px}
    \caption{\textbf{Performance of Basic Operations.}}
    \label{fig:eval-op-overhead}
    \vspace{-5px}
    \end{figure}

\begin{figure*}[t]
\centering
\begin{minipage}[h]{1\linewidth}
    \centering
    \includegraphics[width=.53\linewidth]{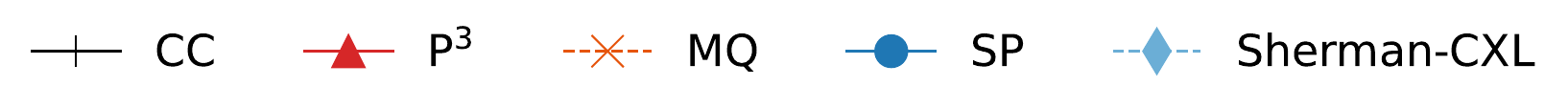}
    \vspace{-6px}
\end{minipage}
\begin{minipage}[t]{1\linewidth}
    \begin{minipage}[t]{.245\linewidth}
        \centering
        \includegraphics[width=1\linewidth]{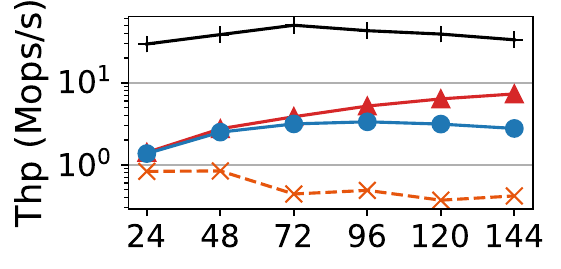}
        \vspace{-20px}
        \subcaption{CLevelHash, A}
        \label{fig:rwtree_performance_fix_db_clevel_workloada_zipfian}
    \end{minipage}
    \begin{minipage}[t]{.245\linewidth}
        \centering
        \includegraphics[width=1\linewidth]{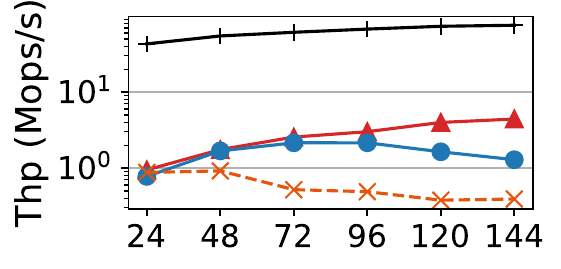}
        \vspace{-20px}
        \subcaption{CLevelHash, B}
        \label{fig:rwtree_performance_fix_db_clevel_workloadb_zipfian}
    \end{minipage}
    \begin{minipage}[t]{.245\linewidth}
        \centering
        \includegraphics[width=1\linewidth]{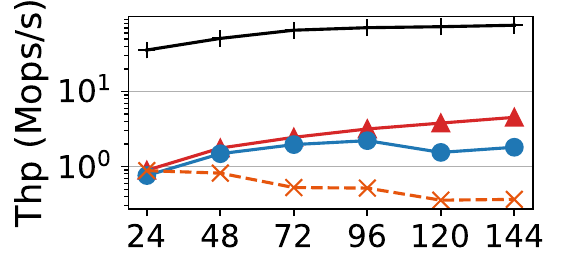}
        \vspace{-20px}
        \subcaption{CLevelHash, C}
        \label{fig:rwtree_performance_fix_db_clevel_workloadc_zipfian}
    \end{minipage}
    \begin{minipage}[t]{.245\linewidth}
        \centering
        \includegraphics[width=1\linewidth]{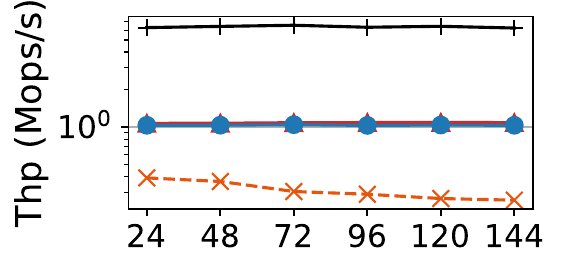}
        \vspace{-20px}
        \subcaption{CLevelHash, Load}
        \label{fig:rwtree_performance_fix_db_clevel_workloadh_zipfian}
    \end{minipage}
\end{minipage}
\begin{minipage}[t]{1\linewidth}
    \begin{minipage}[t]{.245\linewidth}
        \centering
        \vspace{-5px}
        \includegraphics[width=1\linewidth]{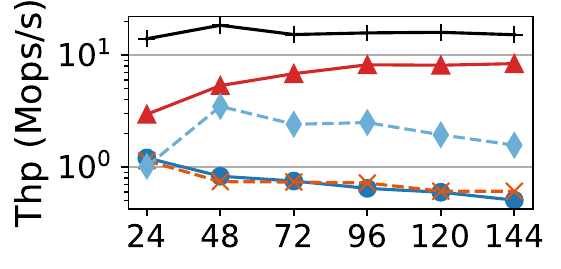}
        \vspace{-20px}
        \subcaption{BwTree, A}
        \label{fig:rwtree_performance_fix_db_ycsb_workloada_zipfian}
    \end{minipage}
    \begin{minipage}[t]{.245\linewidth}
        \centering
        \vspace{-5px}
        \includegraphics[width=1\linewidth]{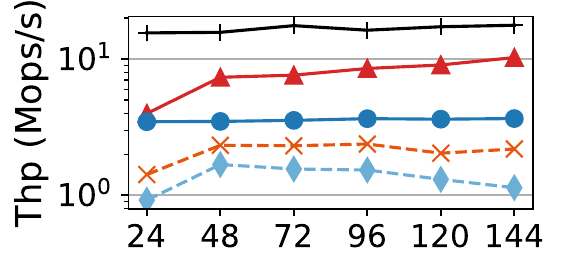}
        \vspace{-20px}
        \subcaption{BwTree, B}
        \label{fig:rwtree_performance_fix_db_ycsb_workloadb_zipfian}
    \end{minipage}
    \begin{minipage}[t]{.245\linewidth}
        \vspace{-5px}
        \centering
        \includegraphics[width=1\linewidth]{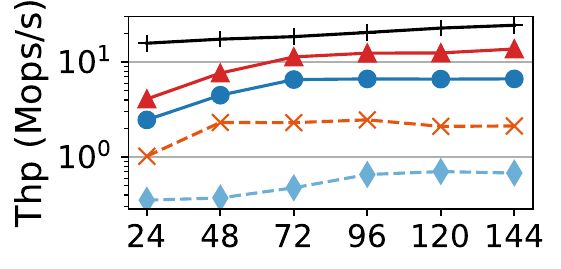}
        \vspace{-20px}
        \subcaption{BwTree, C}
        \label{fig:rwtree_performance_fix_db_ycsb_workloadc_zipfian}
    \end{minipage}
    \begin{minipage}[t]{.245\linewidth}
        \vspace{-5px}
        \centering
        \includegraphics[width=1\linewidth]{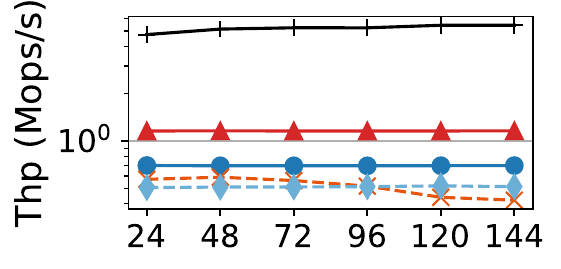}
        \vspace{-20px}
        \subcaption{BwTree, Load}
        \label{fig:rwtree_performance_fix_db_ycsb_workloadh_zipfian}
    \end{minipage}
\end{minipage}

\vspace{-5px}
\caption{\textbf{Performance and scalability of \syshash and \sysbwtree on YCSB workloads.
\vspace{-5px}
}}
\label{fig:eval-ycsb}
\vspace{-5px}
\end{figure*}

\begin{figure}[ht]
    \centering
    \begin{minipage}[t]{1\linewidth}
        \includegraphics[width=1\linewidth]{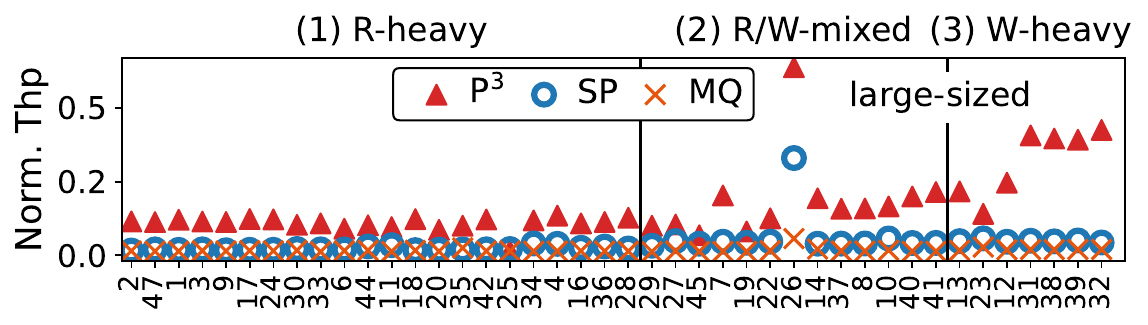}
        \vspace{-20px}
        \subcaption{CLevelHash}
        \label{fig:eval-twitter-clevel}
        \vspace{-3px}
    \end{minipage}
    \begin{minipage}[t]{1\linewidth}
        \includegraphics[width=1\linewidth]{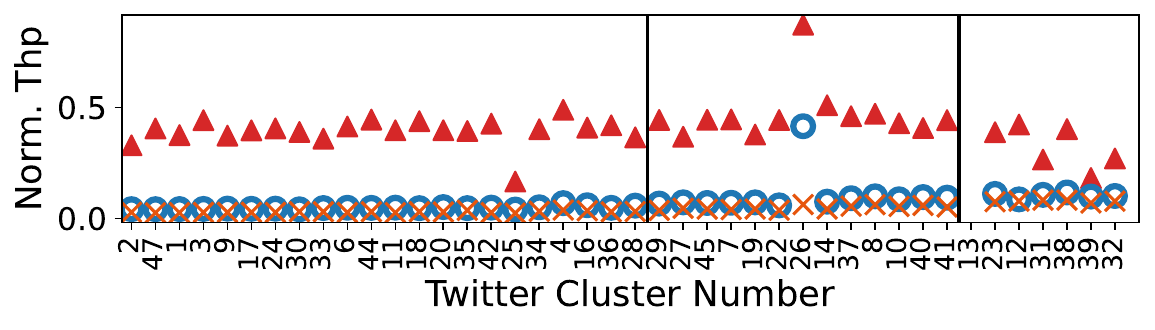}
        \vspace{-20px}
        \subcaption{BwTree}
        \label{fig:eval-twitter-bwtree}
    \end{minipage}  
    \vspace{-13px}
    \caption{\textbf{Performance of \syshash and \sysbwtree on real-world Twitter~\cite{Yang2020TwitterTrace} traces.}}
    \vspace{-5px}
    \label{fig:eval-twitter}
    \end{figure}

\noindent\textbf{Performance of basic operations and \pcas simulation.}
\label{sec:eval-op-overhead}
\autoref{fig:eval-op-overhead} shows the basic performance of our PCC platform.
We use remote CXL memory (CXL-R), which has average latency of 383\,ns and bandwidth of 0.28\,GB/s (tested with MLC~\cite{mlc}).
Due to the lack of hardware support for cross-host \texttt{pCAS}, we simulate \pcas by manually adding latency to standard \texttt{CAS} operation on CXL memory. 
To add latency, each \texttt{pCAS} operation is assigned to one of 4096 queues based on a hash of its target address, and latency is added via a while-loop according to the current queue depth. 
After the latency has been injected, the \texttt{pCAS} operation is removed from the queue.
With this simulation, the bandwidth of \texttt{pCAS} is also controlled since our simulation limits the number of concurrently processed operations.
\texttt{pCAS} exhibits average latencies of 474\,ns and 9\,$\mu$s with 1 and 64 threads, respectively.

\subsection{Performance and Scalability}
\label{sec:eval-performance}

This section evaluates the performance and scalability of \syshash and \sysbwtree against YCSB zipfian workloads~\cite{YCSB2010} and real-world Twitter traces~\cite{Yang2020TwitterTrace}.

\subsubsection{YCSB}

\noindent\textbf{CLevelHash.}
\autoref{fig:eval-ycsb}(a)--(d) shows that \syshash achieves 6--21\% performance of \emph{CC-ClevelHash} on DRAM.
The performance downgrade is caused by several unavoidable \pload, i.e., \pload of ClevelHash's local \emph{ctx\_ptr} and each pointer to the key/value data.
\syshash achieves 10--17$\times$ the throughput of the message passing-based ClevelHash (\emph{MQ-ClevelHash}) and achieves 2.5--3.4$\times$ the throughput of the converted but not optimized \pcchash (\emph{SP-ClevelHash}).
The performance of \emph{MQ-ClevelHash} is limited by message passing overhead, while the performance of \emph{SP-ClevelHash} is bounded by the concurrent \pload of ClevelHash's \emph{global ctx\_ptr}.
For Load operations, the performance of \syshash and \emph{SP-ClevelHash} is comparable, as both are primarily bounded by insert logic rather than the concurrent load on the \emph{global ctx\_ptr}.

\noindent\textbf{BwTree.}
\autoref{fig:eval-ycsb}(e)--(h) shows that \sysbwtree achieves 21\% to 58\% performance of \emph{CC-BwTree} on DRAM.
\sysbwtree achieves 2.8--14$\times$, 1.6--17$\times$ the throughput of \emph{MQ-BwTree}, \emph{SP-BwTree}, respectively.
\emph{SP-BwTree} is bounded by (1) the concurrent \pload of BwTree's global context and (2) the \pload of each node visited during traversing the tree.
In contrast, \sysbwtree only requires one \pload of the leaf-node pointer in the fast path.
Besides, \sysbwtree achieves 2.3--20$\times$ the throughput of \emph{Sherman-CXL}, as Sherman's performance is bounded by the overhead of looking up the client-side index first, and acquiring two-level locks.

\subsubsection[short]{Real-world Twitter~\cite{Yang2020TwitterTrace} traces}

\autoref{fig:eval-twitter}(a)--(b) shows the throughputs normalized to \emph{CC-BwTree} and \emph{CC-CLevelHash} when running Twitter traces.

\noindent\textbf{CLevelHash.}
\autoref{fig:eval-twitter-clevel} shows that \syshash achieves 1.6--64\% (average 18\%) of the throughput of \emph{CC-CLevelHash}.
The write-heavy workloads shows better performance since insert requires fewer \pload operations compared to lookup since it only needs to find the first empty slot.
Note that in cluster \#26, the \syshash\ index shows relatively high values because the value size in this trace is particularly large. Consequently, most of the index operation time is spent on loading values.
Besides, \syshash achieves 0.73--10$\times$ (average 4.8$\times$) of the throughput of \emph{SP-CLevelHash} and 1.1--23$\times$ (average 10$\times$) of the throughput of \emph{MQ-CLevelHash}.
\emph{SP-CLevelHash} and \emph{MQ-CLevelHash} are bounded by \pload and message passing, respectively.

\noindent\textbf{BwTree.}
\autoref{fig:eval-twitter-bwtree} shows that \sysbwtree achieves 17--87\% (average 40\%) of the throughput of \emph{CC-BwTree} across all Twitter traces.
Besides, \sysbwtree achieves 1.8--10$\times$ (average 6.3$\times$) and 2.6--16$\times$ (average 10$\times$) of the throughput of \emph{SP-BwTree} and \emph{MQ-BwTree}, respectively.
\sysbwtree achieves better performance compared with \emph{SP-BwTree} under read-heavy workloads since it uses speculative reading.

\subsection{Factor Analysis}
\label{sec:eval-factor-analysis}

\begin{figure}
\centering
\begin{minipage}[t]{0.39\linewidth}
    \centering
    \includegraphics[width=1\linewidth]{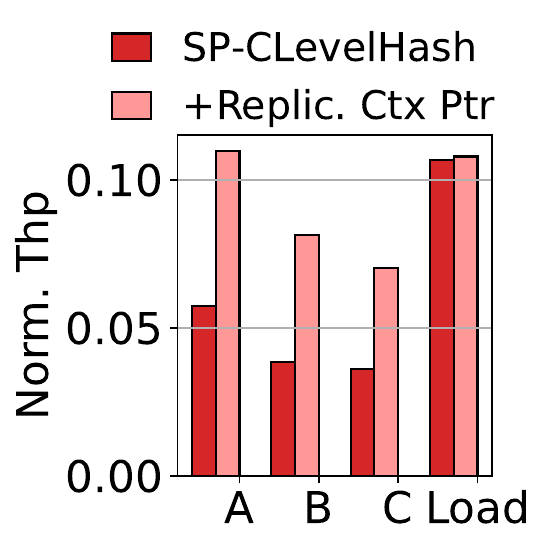}
    \vspace{-22pt}
    \subcaption{CLevelHash}
    \label{fig:eval-factor-analysis-clevel}
\end{minipage}
\begin{minipage}[t]{0.6\linewidth}
    \includegraphics[width=1\linewidth]{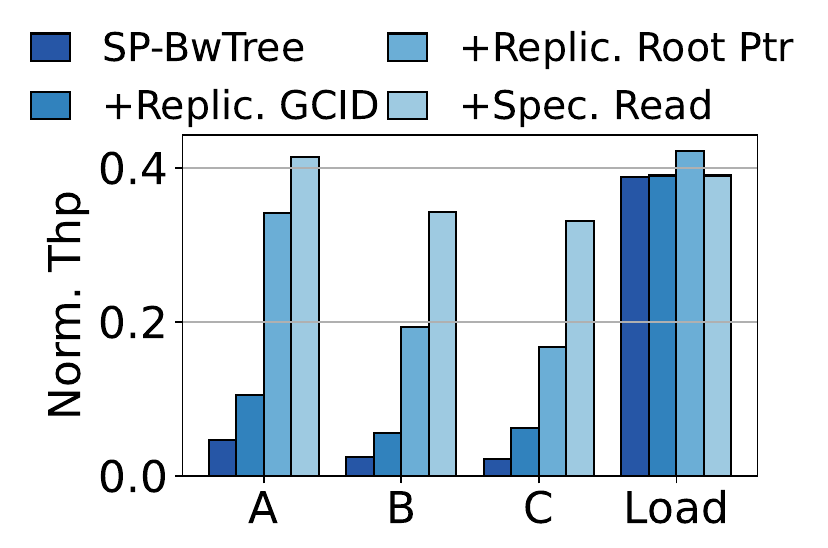}
    \vspace{-22pt}
    \subcaption{BwTree}
    \label{fig:eval-factor-analysis-bwtree}
\end{minipage}
\vspace{-15pt}
\caption{\textbf{Factor analysis.} Performance improvements of each technique, normalized to CC-BwTree and CC-CLevelHash.}
\vspace{-5px}
\label{fig:eval-factor-analysis}
\end{figure}

We conducted factor analysis of \syshash and \sysbwtree with different techniques enabled.
All tests are conducted on 144 threads and the throughputs are normalized to the CC-CLevelHash and CC-BwTree accordingly.

\noindent\textbf{CLevelHash.}
As shown in \autoref{fig:eval-factor-analysis-clevel}, compared not optimized \pcchash,
replicating the \emph{ctx\_ptr} (labeled as \emph{+ Replic. Ctx Ptr}) improves the throughput of workload A, B and C by 91\%, 111\% and 95\% respectively.

\noindent\textbf{BwTree.}
\autoref{fig:eval-factor-analysis-bwtree} shows performance of \pccbwtree after applying three techniques one by one. 
By replicating global GC epoch (\emph{+ Replic. GCID}), the throughput is improved by 125\% to 235\%.
By replicating pointer to root node (\emph{+ Replic. Root Node}), the throughput further improves by 32\% to 225\%.
Speculatively reading (\emph{+ Spec. Read}) improves read-heavy workloads by 78\% (B) and 98\% (C), and improves read-write-mixed workload by 21\% (A), since frequently updated data leads to retry of the operation, causing overhead.

\noindent\textbf{\sysbwtree's speculative reading (G3).}
\begin{table}[t]
    \centering
    \caption{\textbf{Improvement and retry ratios of \guideline-BwTree's speculative reading.}
    Detailed read ratio is shown in \autoref{fig:obs-read-heavy}.
    }
    \label{tab:speculative-read-analysis}
    \vspace{-5px}
    \resizebox{\columnwidth}{!}{
    \begin{tabular}{lcc}
    \hline
    & (1) + (2) Read > 50\% & (3) Read < 50\%\\
    \hline
    Improve & 1\%--120\%$\times$ (Avg.=61\%) & -52\%--129\%$\times$ (Avg.=32\%) \\
    Retry Ratio & 0.0006\%--6.7\% (0.87\%) & 13.7\%--99.7\% (83.4\%) \\
    \hline
    \end{tabular}
    }
    \vspace{-5pt}
\end{table}
\autoref{tab:speculative-read-analysis} shows the improvements (over BwTree applying G2) and retry ratios of \sysbwtree's speculative reading on Twitter traces.
For read-heavy workloads (read ratio > 50\%), speculative reading improves throughput by 1\%--120\% (average 61\%), primarily due to the  reduced \pload and low retry ratios (average 0.87\%).
In contrast, for write-heavy workloads (read ratio < 50\%), speculative reading yields an average throughput improvement of 32\%, but in the worst case, throughput decreases by up to 52\%.
This performance degradation results from increased retry ratios (13.7\%--99.7\%, average 83.4\%) caused by frequent data updates.

\subsection{Real Application: Ray}
\label{sec:eval-real-app}

We integrated {\sysbwtree} into Ray~\cite{ray2018} (version 2.37).
Ray uses Plasma~\cite{plasma} as its distributed object store for inter-worker data transfer across hosts. 
We replaced Plasma with \emph{\guideline-Store}, a cross-host object store based on \sysbwtree.
\guideline-Store enables direct object access in a shared-everything architecture, eliminating data copying, serialization, and network overhead.
Additionally, we developed \emph{Plasma-SHM} by porting Plasma to the CXL platform.
\emph{Plasma-SHM} retains Plasma's original message-passing control plane, but replaces data copying with pass-by-reference transfer (i.e., transferring pointers instead of actual data).

\begin{figure}
\begin{minipage}[t]{0.68\linewidth}
    \centering
    \includegraphics[width=0.46\linewidth]{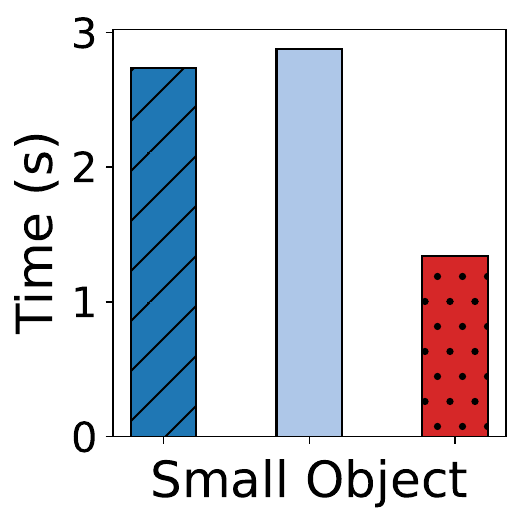}
    \includegraphics[width=0.45\linewidth]{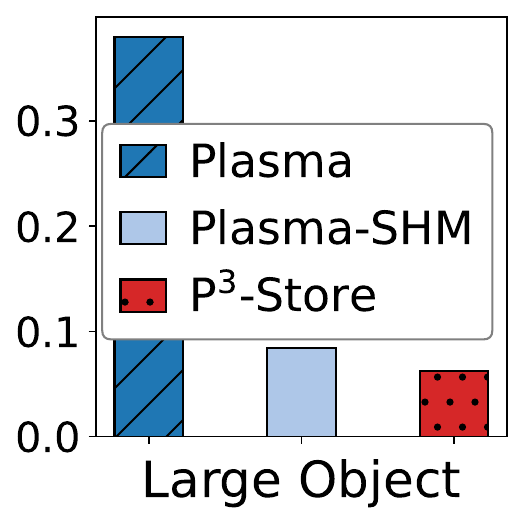}
    \vspace{-8pt}
    \subcaption{Object Store}
    \label{fig:ray-objstore}
\end{minipage}
\begin{minipage}[t]{0.3\linewidth}
    \centering
    \includegraphics[width=1\linewidth]{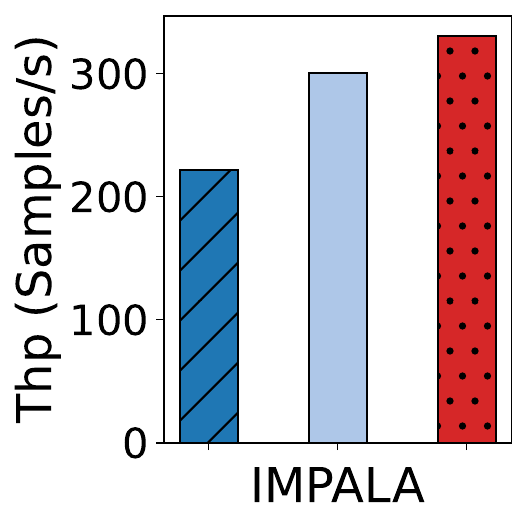}
    \vspace{-20pt}
    \subcaption{RL (IMPALA)}
    \label{fig:ray-rllib}
\end{minipage}
\vspace{-5pt}
\caption{\textbf{Ray Performance with \guideline-Store integrated.}}
\vspace{-5pt}
\label{fig:eval-ray}
\end{figure}

\noindent\textbf{Object store.}
First, we evaluated \guideline-Store 
to show its improvements in metadata and data transfer.
We used two object types: (1) 1,000 small objects (128 KiB each) distributed across 8 hosts, and (2) a single large object (125 MiB) broadcast to all 8 hosts.
As shown in \autoref{fig:ray-objstore}, \guideline-Store reduces transfer time for small objects by 51\% and 53\% compared to \emph{Plasma} and \emph{Plasma-SHM}, respectively, by minimizing metadata and data transfer overheads.
For large objects,  \guideline-Store completes the transfer 80\% faster than \emph{Plasma}.
\emph{Plasma-SHM} also significantly improves upon \emph{Plasma} by reducing data transfer, the primary bottleneck for large object transfers.

\noindent\textbf{Reinforcement learning (RL).}
We use RLlib~\cite{liang2017rllib}, a popular RL library for Ray, to run IMPALA~\cite{impala}, one of the most prevalent RL algorithms.
IMPALA employs a centralized trainer that periodically broadcasts the policy to a set of workers and collects rollouts generated by these workers to update the model.
We emulate four hosts, each equipped with eight cores, and set the batch size to five.
As shown in \autoref{fig:ray-rllib},  \guideline-Store increases the throughput of IMPALA by 49\% and 10\% compared to \emph{Plasma} and \emph{Plasma-SHM}, respectively. 
This improvement is primarily attributed to \guideline-Store's ability to reduce both data and metadata transfer overhead.

\section{Discussion}%
\label{sec:discussion}

\noindent
\textbf{Assumptions on failures.}
Currently, we rely on the memory allocator to reclaim memory after failures.
Specifically,  we use CXL-SHM~\cite{zhang2023partial}, a user-space CXL memory allocator designed to avoid memory leaks.
Indexes often use pre-allocations, so CXL-SHM's overhead is negligible.
Besides, we only consider the failure of the processing hosts but not shared memory pool failures.

\noindent
\textbf{Detectable recovery.}
Another common crash consistency assumption is \emph{detectable recovery}~\cite{friedman2018DL} (i.e., detect whether an operation was completed and if so, tell what its response was, thus preventing undesirable outcomes of re-executing completed operations).
\ds does not require detecting the completion of the operations, as index operations are idempotent and re-executing always gets the same outcomes.

\noindent
\textbf{Adapting to other PCC models.}
Recent studies~\cite{wang2025enabling,tang2024Exploring} suggest that a small-region memory can support hardware cache-coherence across multiple hosts.
However, this solution does not scale as the number of hosts increases according to existing reports~\cite{wang2025enabling} and our discussions with CXL vendors.
Also, our observations and techniques can be applied to such a model: we can place sync-data on this small-region cache-coherence memory and leverage hardware cache-coherence to synchronize them more efficiently.

\section{Related Work}%
\label{sec:related}

\noindent
\textbf{Software-Managed Cache Coherence.}
Previous software-managed cache coherence proposals have demonstrated that comparable or even superior performance can be achieved with minimal or no hardware modifications~\cite{fensch2008OSbased,kelm2009Rigela,zhou2010Case}.
However, these models are typically designed for applications characterized by little data sharing, while we target indexes, which involve extensive data sharing.
The Bounded Incoherence~\cite{ren2020Bounded} aims to support software cache coherence for shared indexes using an RCU~\cite{Desnoyers2012RCU}-like programming model; however, it necessitates non-trivial modifications and ensures release consistency.
In contrast, we provide a simple and efficient programming model with linearizability.

\noindent
\textbf{Designing Indexes for New Platforms.}
Designing correct and efficient indexes has been a long-standing challenge.
With the advent of new hardware platforms, many works have focused on designing indexes for them.
We draw on insights from persistent indexes~\cite{Lee2019RecipeCC, Ramanathan2021TIPSMV} for making indexes durable linearizability, and DRAM-based disaggregated platforms~\cite{sherman2021,luo2023SMART,zuo2021Onesided} for using local cache to improve efficiency. %
\section{Conclusion}%
\label{sec:concl}

This paper focuses on making indexes correct and efficient on PCC platforms. 
We introduce \glshort, i.e., the \rules to convert indexes designed for cache-coherence platforms to run on PCC platforms correctly, and {\gl}s to make the converted indexes efficient.
We evaluate indexes converted with \glshort on YCSB and Twitter traces, and integrate them into a real-world application (Ray) to show \glshort's benefits.

\clearpage
\bibliographystyle{plain}
\bibliography{paper,website}

\clearpage
\appendix

\section{Details and Correctness of Converted PCCIndex}
\label{sec:appendix-code-indexes}

This section shows detailed code of how we convert a OLC-based and lock-free CCIndex to PCCIndex.

\subsection{OLC-based Index}%

Optimistic lock decoupling (OLC) is a technique that is used by a wide range of indexes (e.g., Masstree, ARTOLC).
We use an OLC-based B-tree index as an example to show how we convert a OLC-based CCIndex to PCCIndex.

\begin{figure}[h!]
    \centering
    \includegraphics[width=1\linewidth]{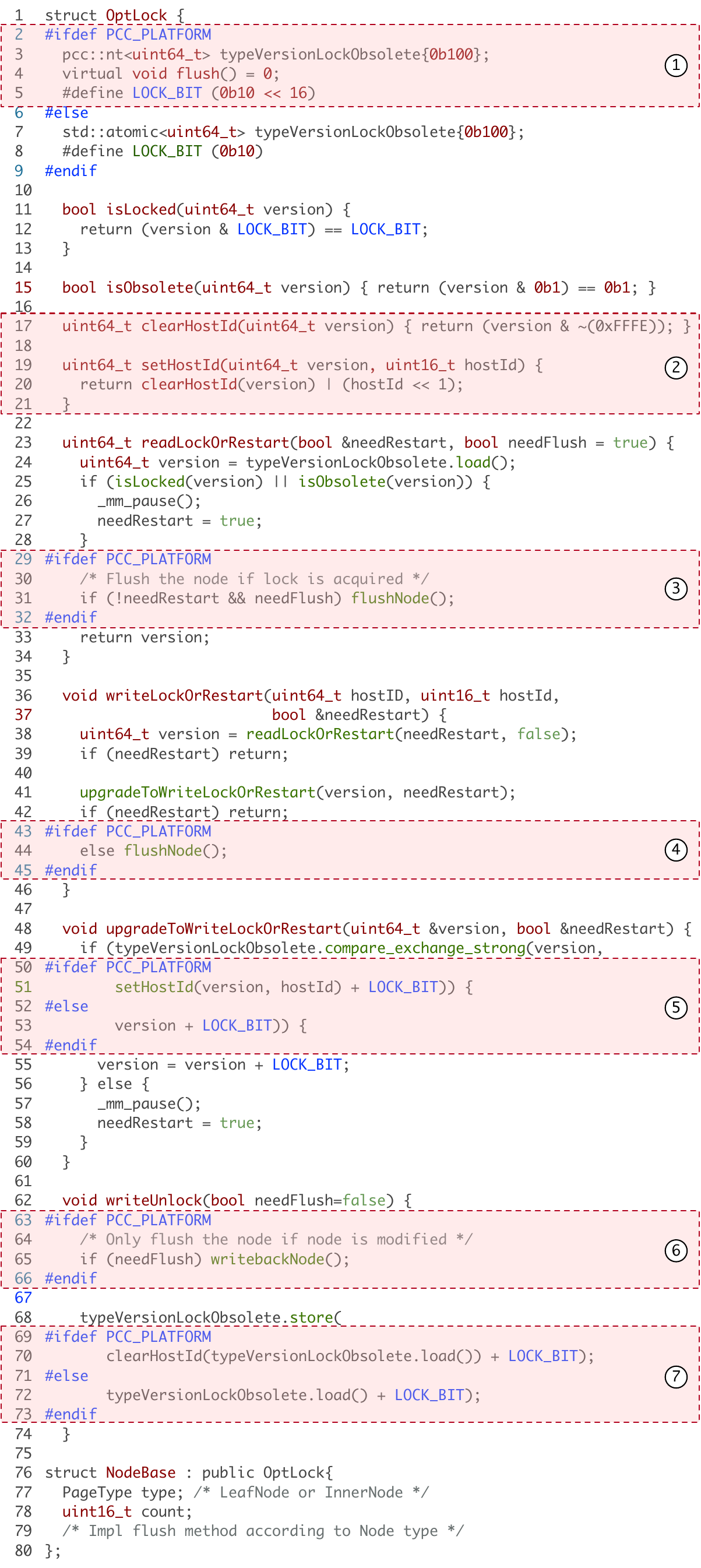}
    \vspace{-10px}
    \caption{C++ Code of lock in Optimistic Lock Decoupling}
    \label{fig:btree-olc}
\end{figure}

\autoref{fig:btree-olc} shows detailed code of OLC-based B-tree index\footnote{The orignial code can be found in \url{https://github.com/cosmoss-jigu/hydralist/blob/master/BTreeOLC/BTreeOLC.h}}, together with what we add to enable consistency on PCC platforms.
We added code is contained in boxes.

\emphbf{Concurrent-safe (R1).}
To ensure concurrent-safe of OLC-based indexes, we replace the \texttt{std::atomic} \emph{version} field in the lock with our \texttt{pcc::nt} pointer (code block \ding{172}). 
The \texttt{pcc::nt} pointer automatically performs \pload/\pstore/\pcas operations, enabling existing codes to easily switch to cache-bypass operations by simply replacing their \texttt{std::atomic} pointer declarations.
For example, the original \texttt{compare\_exchange\_strong} operation of the \emph{version} field is atomically replaced by \pcas.
Thereby, we ensure the consistency of the sync-data (i.e., the lock) on PCC platforms.
Besides, we also need to write-back the protected data (i.e., the node's data) to memory.
We add a \texttt{flushNode} function, and implement it according to the node type.
Specifically, we use \texttt{clflush} + \texttt{mfence} to invalidate the node's cacheline and use \texttt{clwb} + \texttt{mfence} to write-back the node's data to memory.
We add the \texttt{flushNode} function after successfully getting the read lock (\ding{174}), get the write lock (\ding{175}).
Besides, we add the \texttt{writebackNode} function after releasing the write lock (\ding{177}).

\emphbf{Durable linearizable (R2.1).}
DL is guaranteed by the cache-bypass update of the \emph{version} field and the flush of node's data.
With \pcas, the linearizability point and the durable point are the same.
Besides, we flush of node's data after each store operation, following the technique developed for persistent memory (PM) indexes~\cite{Lee2019RecipeCC}.

\emphbf{Failure isolated (R2.2).}
We use a \emph{recoverable lock} to ensure failure isolation.
Generally, the \emph{recoverable lock} encoded the ID of the host in the lock; thus, when we see a lock is acquired by a host, we can know welther the the lock's owner is still alive.
The liveness of each host is checked by the controller following the method proposed in Lupin~\cite{zhu2024Lupin}.

Specifically, the \emph{recoverable lock} encoded a 16-bit \emph{hostId} in the 1-16 bit of the 64-bit \emph{version} field.
The 0 bit of \emph{version} is used to indicate the whether the lock is deleted, and the 17-bit is the lock bit, which indicates whether the lock is acquired by any writer.
When trying to get the write lock, the writer also need to encode its \emph{hostId} into the \emph{version} field with \texttt{setHostId} (code block \ding{176}).
When releasing the write lock, the writer also need to clear the \emph{hostId} atomically with clearing the lock bit (code block \ding{178}).

\subsection{Lock-free Index}%

\begin{figure}[h!]
    \centering
    \includegraphics[width=1\linewidth]{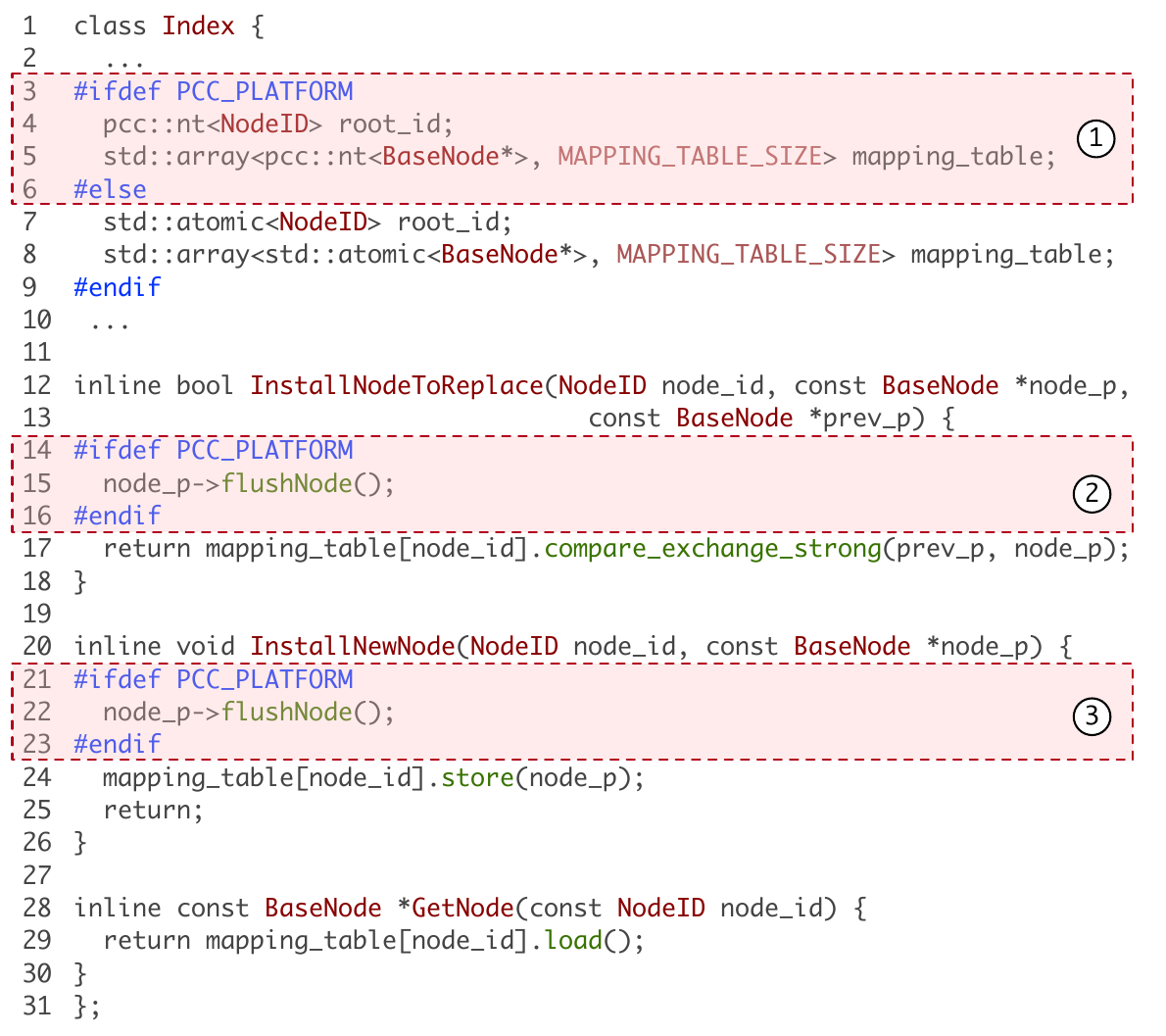}
    \vspace{-10px}
    \caption{C++ Code of Lock-free BwTree}
    \label{fig:lockfree-index}
\end{figure}

We use BwTree as an example to show how we extend lock-free index to work correctly on PCC platforms.
The C++ code and our modification is shown in \autoref{fig:lockfree-index}.

\emphbf{Concurrent-safe (R1).}
To ensure concurrent-safe of lock-free indexes, we replace the \texttt{std::atomic} \emph{root\_id} and pointer in the mapping table (\emph{mapping\_table[node\_id]}) with our \texttt{pcc::nt} pointer (code block \ding{172}).
With \texttt{pcc::nt}, installing nodes or accessing nodes from the mapping table will be replaced by cache-bypass operations.
Besides, we also need to flush the newly allocated node with \texttt{clwb} + \texttt{mfence} as shown in code block \ding{173} and \ding{174}.

\emphbf{Durable linearizable (R2.1) and Failure isolated (R2.2).}
RECIPE already proposes a converted BwTree that is durable linearizable, and we follow their method.
Besides, since BwTree is a lock-free index, failure isolated is already guaranteed since no one will hold the lock.

\section{Details of Replicating for DGC}%
\label{sec:appendix-G2-gc-blocking}

We give a concrete example to show inconsistency caused by replicating DGC's global epoch, and shown why our method can fix this inconsistency.

\begin{figure}[h!]
    \centering
    \includegraphics[width=1\linewidth]{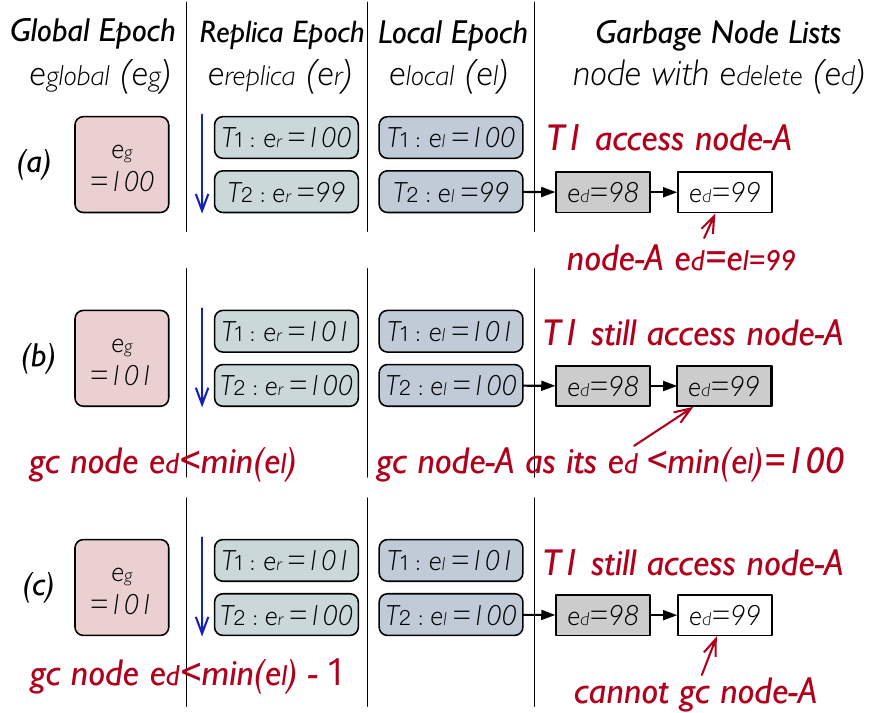}
    \vspace{-10px}
    \caption{Example of inconsistency caused by replicating DGC's global epoch}
    \label{fig:gc-blocking}
\end{figure}

\autoref{fig:gc-blocking} shows an example of inconsistency caused by replicating DGC's global epoch.
In (a), after $T_{gc}$ increments $e_{g}$ to 100, it updates $T_1$'s $e_{r}$ to 100.
$T_{1}$ then executes $e_{l}:=e_r$ (the value is 100) and starts accessing a node (termed node-A).
Meanwhile, $T_{2}$ executes $e_{l} := e_r$ (the value is 99 since $T_{gc}$ has not updated $T_2$'s $e_r$) and removes node-A by appending it to the garbage node lists with $e_{d}$=99.

While $T_1$ is still accessing node-A, $T_2$ can garbage collect node-A once $T_2$'s $e_l$ becomes larger than 99 (in \autoref{fig:gc-blocking} (b)), causing use-after-free issues.

The root cause is that not all $e_r$ are updated atomically and in the above example, $T_2$'s $e_r$ can be smaller than $T_1$'s $e_r$ by one.
After fix the inconsistency by altering the removal condition from ``<$min(e_{l})$'' to ``< $min(e_{l})-1$'' (step \darkred{\ding{174}} in \autoref{fig:tech-gc}), this mistakenly removed node-A will not be removed again.
As shown in \autoref{fig:gc-blocking} (c), node-A's $e_{d}$ is 99, which does not satisfy the removal condition as $min(e_{l})-1$ is 99, so node-A will not be removed again.

\end{document}